# Magnetic Penetration Depth in Unconventional Superconductors

**Ruslan Prozorov**

Ames Laboratory and Department of Physics & Astronomy, Iowa State University, Ames, Iowa 50011, U. S. A.

**Russell W. Giannetta**

Department of Physics, University of Illinois at Urbana-Champaign, 1110 W. Green Street, Urbana, IL 61801, U. S. A.

**Abstract**. This topical review summarizes various features of magnetic penetration depth in unconventional superconductors. Precise measurements of the penetration depth as a function of temperature, magnetic field and crystal orientation can provide detailed information about the pairing state. Examples are given of unconventional pairing in hole- and electron-doped cuprates, organic and heavy fermion superconductors. The ability to apply an external magnetic field adds a new dimension to penetration depth measurements. We discuss how field dependent measurements can be used to study surface Andreev bound states, nonlinear Meissner effects, magnetic impurities, magnetic ordering, proximity effects and vortex motion. We also discuss how penetration depth measurements as a function of orientation can be used to explore superconductors with more than one gap and with anisotropic gaps. Details relevant to the analysis of penetration depth data in anisotropic samples are also discussed.

## 1. Introduction

The early suggestion that high temperature superconductors might exhibit unconventional, d-wave pairing, [24, 11, 154, 8] has lead to a wide variety of new experimental probes with sensitivity sufficient to test this hypothesis. The pioneering work of Hardy and coworkers demonstrated that high resolution measurements of the London penetration depth could detect the presence of nodal quasiparticles characteristic of a d-wave pairing state [87]. Since that time, a large number of new superconductors have been discovered, many of which exhibit nontrivial departures from BCS behavior. As we show in this review, penetration depth measurements can be used to examine not only nodal quasiparticles but two-gap superconductivity, anisotropy of the energy gap, Andreev surface states, nonlinear Meissner effect, interplane transport, proximity coupled diamagnetism, vortex matter and the coexistence of magnetism and superconductivity, to name several problems of current interest.

Strictly speaking, the term "unconventional" superconductivity refers to the case where the symmetry of the pairing state (defined later) is lower than that of the crystal lattice. [11, 7] The case familiar to most researchers is the famous d-wave pairing state. Although much of the article will focus on superconductors of this type, many interesting new materials have been discovered for which penetration depth techniques, originally developed to search for unconventional pairing, have become invaluable. It is also important to demonstrate how effectively penetration depth measurements can distinguish between an unconventional superconductor and, for example, one with an anisotropic energy gap. We will therefore extend our discussion to include some of these materials.

This article is not intended as a comprehensive review of all experimental methods designed to measure penetration depth. Instead, we focus on the radio frequency oscillator methods developed in our own laboratories, largely to demonstrate what can be observed with this very high resolution technique. We have also provided a discussion of various technical aspects of penetration depth measurements related to sample shape and anisotropy. Some general reviews that also discuss penetration depth in unconventional superconductors can be found in the reference section. Since space is limited, we will assume familiarity with the basic concepts of the BCS model [17] and refer the reader to any standard text for a more detailed discussion. For some further reading, the reader is encouraged to explore the following articles: [61, 9, 7, 44, 43, 111, 140, 201, 34, 206, 207, 88, 58, 103, 20]

1.1. Pairing states

We begin by defining the term "pairing state", our discussion closely following that of Annett *et al.* [11, 7]. Pairing will mean a non-zero expectation value of the "gap matrix",

$$\Delta_{\alpha\beta}(\mathbf{k}) \propto \langle c_{\mathbf{k}\alpha} c_{-\mathbf{k}\beta} \rangle \tag{1}$$

where α and β are spin indices, $\mathbf{k}$ is the wavevector and $c_{\mathbf{k}\alpha}$ is a destruction operator for a charge carrier in state $|\mathbf{k},\alpha\rangle$. A general form for the gap matrix is,

$$\Delta_{\alpha\beta}(\mathbf{k}) = i\left[\Delta(\mathbf{k}) + \boldsymbol{\sigma}\cdot\mathbf{d}(\mathbf{k})\right]\sigma_y \tag{2}$$

where $\sigma_i$ are Pauli matrices. Since the wave function for two fermions must be odd under interchange, $\Delta(\mathbf{k})$ must be even and the vector $\mathbf{d}(\mathbf{k})$ must be an odd function of $\mathbf{k}$. Spin singlet states correspond to $\Delta(\mathbf{k})\neq 0, \mathbf{d}(\mathbf{k})=0$. These can be thought of as having even orbital angular momentum quantum numbers. (s,d,g-wave etc.) Most superconductors in nature fall into this category. Spin triplet states correspond to $\Delta(\mathbf{k})=0, \mathbf{d}(\mathbf{k})\neq 0$ and have odd orbital angular momentum. The best known examples of spin triplet pairing are the p-wave states of superfluid $^3$He. [128]. Higher angular momentum states have been proposed. For example an f-wave has been suggested for $Sr_2RuO_4$. [212, 155] It is also possible to have mixed symmetries such as $s+id$, $s+d$ and for lattices that do not possess inversion symmetry, one can in principle even have singlet and triplet combinations. In most of the literature the term "unconventional" superconductivity refers to the situation where the pairing state has a lower symmetry than the point group of the underlying crystal lattice. For example, the now famous $d_{x^2-y^2}$ state relevant to high temperature superconductors can be represented by a gap function of the form $\Delta(\mathbf{k}) \propto (k_x^2 - k_y^2)$ which changes sign and is therefore unconventional. A gap function which has nodes in the same place as the $d_{x^2-y^2}$ state but is everywhere positive (e.g $\Delta(\mathbf{k}) \propto |k_x^2 - k_y^2|$) would be considered conventional. Since most of the examples in this paper correspond to spin singlets, we will deal with $\Delta(\mathbf{k})$ for the most part and refer to it as the "gap function".

1.2. Isotropic London electrodynamics

The notion of a characteristic length for magnetic field penetration into a superconductor was established soon after the discovery of the magnetic field expulsion in tin and lead by W. Meissner and R. Ochsenfeld in 1933. [152, 153] C. J. Gorter and H. Casimir (GC) introduced a two-fluid model of superconductivity in 1934. [79, 80, 78] Analysis of the specific heat and critical field data, prompted GC to suggest an empirical form for the temperature dependence of the density of superconducting electrons, $n_s = n(1-t^4)$, where $t = T/T_c$ and $n$ is the total density of conduction electrons. In 1935 F. and H. London [134] introduced a phenomenological model of superconductivity in which the magnetic field inside a superconductor **B** obeys the equation,

$$\nabla^2 \mathbf{B} = \frac{\mathbf{B}}{\lambda_L^2} \tag{3}$$

where $\lambda_L$, known as the London penetration depth, is a material-dependent characteristic length scale given by

$$\lambda_L^2 = \frac{mc^2}{4\pi n_s e^2} \tag{4}$$

Combining this definition with the GC form for the density of superconducting electrons results in a temperature dependent penetration depth,

$$\lambda(T) = \frac{\lambda(0)}{\sqrt{1-t^4}} \tag{5}$$

Although Eq.(5) has no microscopic justification, at low temperatures it takes the form $\lambda(T) \approx \lambda(0)(1 + t^4/2 + O(t^8))$ which is nearly indistinguishable from the exponential behavior $\lambda(T) \sim T^{-1/2} \exp(-\Delta/T)$ predicted by BCS theory. [17] The GC expression has often been used to provide a quick estimate of $\lambda(0)$, a quantity that is generally very difficult to determine. Often, the results are quite reasonable.[143] Some successful attempts were made to generalize the GC approximation to better fit the results of the measurements. [129]. Fitting to the microscopic BCS calculations, Eq.(7), produces good fits using $\lambda(T) = \lambda(0)/\sqrt{1-t^p}$ with $p = 2$ for s-wave and $p = 4/3$ for d-wave superconductors.

## 2. Semiclassical approach to the superfluid density

We will briefly describe the main results of a semiclassical model for the penetration depth given by B. S. Chandrasekhar and D. Einzel [48]. Given a Fermi surface and a gap function, this approach provides a general method for calculating all three spatial components of the penetration depth. It is limited to purely coherent electronic transport and does not include the effects of scattering. We restrict ourselves to singlet pairing states.

In the London limit, the supercurrent $\mathbf{j}(\mathbf{r})$ is related locally to the vector potential $\mathbf{A}(\mathbf{r})$ through a tensor equation,

$$\mathbf{j} = -\mathbb{R}\mathbf{A} \tag{6}$$

The (symmetric) response tensor is given by,

$$\mathbb{R}_{ij} = \frac{e^2}{4\pi^3 \hbar c} \oint_{FS} dS_\kappa \left[ \frac{v_F^i v_F^j}{|\mathbf{v}_F|} \left( 1 + 2\int_{\Delta(\kappa)}^\infty \frac{\partial f(E)}{\partial E} \frac{N(E)}{N(0)} dE \right) \right] \tag{7}$$

Here $f$ is the Fermi function and $E = \sqrt{\varepsilon^2 + \Delta^2}$ is the quasiparticle energy. (The normal metal band energy $\varepsilon$ is measured from the Fermi level.) $N(E)/N(0) = E/\sqrt{E^2 - \Delta(\mathbf{k})^2}$ is the density of states normalized to its value at the Fermi level in the normal state and $v_F^i$ are the components of Fermi velocity, $\mathbf{v}_F$. The average is taken over the Fermi surface, with the a **k**-dependent superconducting gap $\Delta(\mathbf{k},T)$. We note that often Fermi surface averaging is used only on the second integral term in Eq. (7). This may lead to significant deviations in calculating the superfluid density if Fermi surface is not spherical. Using a coordinate system defined by the principal axes of $\mathbb{R}$ one has,

$$\mathbf{j}(\mathbf{r}) = \frac{c}{4\pi} \frac{\mathbf{A}(\mathbf{r})}{\lambda_{ii}^2} \tag{8}$$

The penetration depths are given by,

$$\lambda_{ii}^2 = \frac{c}{4\pi \mathbb{R}_{ii}} \tag{9}$$

and the superfluid density components are given by,

$$n_{ii}(T) = \frac{cm_{ii}}{e^2} \mathbb{R}_{ii}(T) \tag{10}$$

with the effective mass defined as

$$m_{ii} = \frac{e^2 n}{c \mathbb{R}_{ii}(0)} \tag{11}$$

As a simple example, for a spherical Fermi surface, one obtains the total electron density, $n_{ii}(T \to 0) = n = k_F^3/3\pi^2$. The *normalized* superfluid density components are given by

$$\rho_{ii}(T) = \frac{n_{ii}(T)}{n} = \frac{\mathbb{R}_{ii}(0)}{\mathbb{R}_{ii}(T)} = \left(\frac{\lambda_{ii}(0)}{\lambda_{ii}(T)}\right)^2 \tag{12}$$

Equation (7) provides the connection between experimentally measured penetration depth and microscopic superconducting state. Without any further calculation, these formulas demonstrate the important point that in the clean limit $\lambda(T=0)$ is simply a band structure property, unrelated to the gap function. For superconductors with sufficiently strong scattering, this statement must be modified. The importance of the gap function becomes evident at non-zero temperatures where it is possible to generate quasiparticle excitations and a paramagnetic current.

In the case of a (2D) cylindrical Fermi surface (often used as simple approximation for the copper oxide superconductors),

$$\rho_{\substack{aa\\bb}} = 1 - \frac{1}{2\pi T} \int_0^{2\pi} \binom{\cos^2(\varphi)}{\sin^2(\varphi)} \int_0^\infty \cosh^{-2}\left(\frac{\sqrt{\varepsilon^2 + \Delta^2(T,\varphi)}}{2T}\right) d\varepsilon d\varphi \tag{13}$$

where $\Delta(\varphi)$ is angle dependent gap function. For a 3D spherical Fermi surface and an anisotropic gap $\Delta(\theta,\varphi)$, Eqs. (7) and (12) become,

$$\rho_{\substack{aa\\bb}} = 1 - \frac{3}{4\pi T} \int_0^1 (1-z^2) \int_0^{2\pi} \binom{\cos^2(\varphi)}{\sin^2(\varphi)} \int_0^\infty \cosh^{-2}\left(\frac{\sqrt{\varepsilon^2 + \Delta^2(T,\theta,\varphi)}}{2T}\right) d\varepsilon d\varphi dz \tag{14}$$

and

$$\rho_c = 1 - \frac{3}{2\pi T} \int_0^1 z^2 \int_0^{2\pi} \cos^2(\varphi) \int_0^\infty \cosh^{-2}\left(\frac{\sqrt{\varepsilon^2 + \Delta^2(T,\theta,\varphi)}}{2T}\right) d\varepsilon d\varphi dz \tag{15}$$

where $z = \cos(\theta)$. We use $k_B = 1$ throughout the manuscript.

For isotropic s-wave pairing both the 2D and 3D expressions give

$$\rho = 1 - \frac{1}{2T}\int_0^\infty \cosh^{-2}\left(\frac{\sqrt{\varepsilon^2 + \Delta^2(T)}}{2T}\right)d\varepsilon \tag{16}$$

2.1. The superconducting gap function

In order to use this model to calculate the superfluid density one requires a form for the gap function. For spin singlet pairing states this takes the form,

$$\Delta(T,\mathbf{k}) = \Delta_0(T)g(\mathbf{k}) \tag{17}$$

$g(\mathbf{k})$ is a dimensionless function of maximum unit magnitude describing the angular variation of the gap on the Fermi surface. $\Delta_0(T)$ carries the temperature dependence and should be determined from the self-consistent gap equation. The latter involves a Fermi surface average of $g(\mathbf{k})$, so in general $\Delta_0(T)$ depends on the pairing symmetry. The question of how to self consistently determine the gap function in the general case where strong coupling corrections are present is outside the scope of this paper. And excellent discussion is given by Carbotte [43]. For simple estimates an approximate form can be used. One of the most useful expressions for $\Delta_0(T)$ was given by Gross *et al.* [85]

$$\Delta_0(T) = \Delta_0(0)\tanh\left(\frac{\pi T_c}{\Delta_0(0)}\sqrt{a\left(\frac{T_c}{T}-1\right)}\right) \tag{18}$$

where $\Delta_0(0)$ is the gap magnitude at zero temperature and $a$ is a parameter dependent upon the particular pairing state (related to the relative jump in a specific heat). Figure 1 shows this function (lines) plotted for four different pairing states. In each case the weak coupling self-consistent gap equation was solved using an appropriate $g(\mathbf{k})$ and plotted as symbols. The fits are very close, with parameters the parameters for **(1)** isotropic s-wave, $g = 1$, $\Delta_0(0) = 1.76T_c$, $a = 1$ **(2)** two-dimensional d-wave: $g = \cos(2\varphi)$, $\Delta_0(0) = 2.14T_c$, $a = 4/3$, **(3)** s+g wave [211], $g = (1 - \sin^4(\theta)\cos(4\varphi))/2$, $\Delta_0(0) = 2.77\,T_c$, $a = 2$ and **(4)** non-monotonic d-wave [147], $g = 1.43\cos(2\varphi) + 0.43\cos(6\varphi)$, $\Delta_0(0) = 1.19T_c$, $a = 0.38$. (Note, that in ARPES measurements the actual superconducting gap can be masked by the pseudogap (e.g., recent reports on BSCCO-2201 [116]. The pseudogap does not contribute to the superfluid density and therefore measurements of the penetration depth are helpful in separating the two gaps).

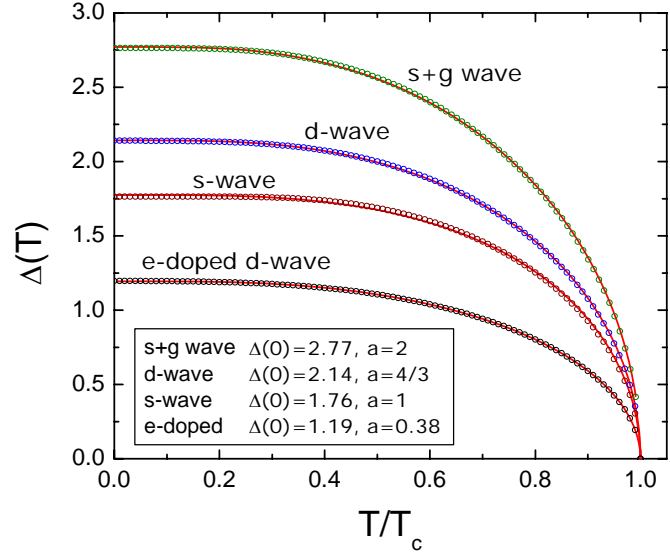

Figure 1. Temperature dependence of the (weak-coupling limit) superconducting gap for s, d, s+g and nonmonotonic d-wave symmetries obtained from self-consistent gap equation (symbols) and fitting to Eq.(18) (solid lines) with fit parameters indicated.

As Figure 1 illustrates, the gap function is very nearly constant below $T/T_c \approx 0.25$ even if nodes are present. It is only in this limit where the temperature dependence of the penetration depth allows one to draw general conclusions about the pairing state. At higher temperatures, the temperature dependence of the gap itself cannot be ignored and strong coupling corrections can change this dependence substantially. In the low temperature (almost constant gap) limit, things are simpler. Defining $\Delta\lambda = \lambda(T) - \lambda(0)$, Eq.(12) can be written as

$$\rho = \left(1 + \frac{\Delta\lambda(T)}{\lambda(0)}\right)^{-2} \approx 1 - 2\frac{\Delta\lambda(T)}{\lambda(0)} \qquad (19)$$

For g = 1 (s-wave) we obtain the standard BCS result in the low temperature limit, [1]

$$\rho \approx 1 - \sqrt{\frac{2\pi\Delta_0}{T}} \exp\left(-\frac{\Delta(0)}{T}\right) \qquad (20)$$

The corresponding penetration depth, $\Delta\lambda/\lambda(0) \approx (1-\rho)/2$ is given by,

$$\frac{\Delta\lambda(T)}{\lambda(0)} \approx \sqrt{\frac{\pi\Delta_0}{2T}} \exp\left(-\frac{\Delta(0)}{T}\right) \qquad (21)$$

The exponentially small value of $\Delta\lambda/\lambda(0)$ justifies the linear approximation of Eq. (19) and shows that $\rho$ and $\lambda$ have the same temperature dependence in this region. However, for a gap function with nodes, $\lambda$ changes much more rapidly with temperature. Higher order terms in Eq. (19) can quickly lead to different $T$ dependences for $\rho$ and $\lambda$, even if the gap does not change with temperature. One must then determine the more fundamental quantity, $\rho$, which requires an additional determination of $\lambda(0)$. For a gap function with nodes, $\rho$ depends upon both the gap topology (point nodes, line nodes, etc.) and the functional behavior of $g(\mathbf{k})$ near the nodes.[85] In principle, there are an infinite number of gap functions consistent with a given pairing symmetry. For the $d_{x^2-y^2}$ symmetry now believed to describe high temperature cuprates, one widely used choice is $\Delta = \Delta_0(0)\cos(2\varphi)$. At low temperatures this leads to a superfluid density varying as,

$$\rho = 1 - \frac{2\ln 2}{\Delta_0(0)} T \qquad (22)$$

Another possible choice with $d_{x^2-y^2}$ symmetry is a two parameter gap function which varies linearly with angle near the nodes and is constant for larger angles:

$$|\Delta(\varphi)| = \begin{cases} \alpha\Delta_0(0)\varphi, & 0 \leq \varphi \leq \alpha^{-1} \\ \Delta_0(0), & \alpha^{-1} \leq \varphi \leq \pi/4 \end{cases} \qquad (23)$$

where $\alpha = \Delta_0^{-1}(0) d|\Delta(\varphi)|/d\varphi \big|_{\varphi \to \varphi_{node}}$ [213]. This form leads to,

$$\frac{\Delta\lambda(T)}{\lambda(0)} \approx \frac{2\ln 2}{\alpha\Delta_0(0)} T \qquad (24)$$

Eq. (24). reduces to Eq. (22) if $\Delta = \Delta(0)\cos(2\varphi)$ is used. In either case, the linear T dependence results from the linear variation (near the nodes) of the density of states, $N(E) \sim E$. An interesting "nonmonotonic" d-wave gap function is discussed in Section 6.2. The importance of low temperature measurements is this direct access to the geometrical properties of the gap function.

## 3. Experimental methods

A large number of experimental techniques have been developed to measure the penetration depth. We will touch briefly upon some of these methods, but will restrict most examples in this paper to data taken using a tunnel diode oscillator technique developed over the past several years at the University of Illinois. [46, 47, 168, 169] This is essentially a cavity perturbation technique, albeit at rf frequencies. In the microwave approach [88, Jacobs, 1995 #218, Mao, 1995 #260] a very high quality factor cavity is typically driven externally while its in-phase and quadrature response are measured. In the tunnel diode method, the "cavity" is simply an inductor, typically copper, with a Q of order 100. The coil forms part of an LC tank circuit that is driven to self-resonate with a negative resistance tunnel diode. Tunnel diode oscillators were used early in low temperature physics, but the potential of this approach for very high resolution measurements was first demonstrated by van de Grift. [60] By placing a superconducting sample inside the coil, changes in the penetration depth, or more precisely its rf magnetic susceptibililty, result in changes of inductance and therefore oscillator frequency. With care, frequency resolution of $10^{-9}$ in a few seconds counting time can be achieved. For the sub-mm sized crystals characteristic of modern superconductivity research, this resolution translates into changes in λ of order 0.5 Å or smaller. In order to isolate variations of the sample temperature from the oscillator, the sample is attached to a sapphire cold finger [196] whose temperature can be varied independently. The cold finger stage can be moved, *in situ*, to determine field and temperature dependent backgrounds and to calibrate the oscillator response. [46] We mention that Signore *et al.* used a tunnel diode oscillator to observe unconventional superconductivity in the heavy fermion superconductor UPt$_3$ [189] More recent variations that provide sample/oscillator thermal isolation have been developed to measure $\lambda(T)$ down to 50 mK. [50, 29, 30, 67]

A great advantage of the tunnel diode method is that the resonator need not be superconducting so that external magnetic fields can be applied to the sample. This feature was exploited early on to search for possible changes in the penetration depth of YBCO with magnetic field [196] and to study vortex motion [101]. Tunnel diode oscillators with suitably designed coils are now used in very large and even pulsed magnetic fields. [54] The relatively low quality factor of the LC combination means that AC excitation fields can be very small (~20 mOe) and thus do not perturb the sample in any significant way. Unless one is very close to $T_c$, the imaginary part of the sample conductance, and thus the penetration depth, dominates the response. The oscillator approach provides tremendous resolution, but one does not have direct access to the dissipative component of the response, as is possible in a driven cavity or mutual inductance technique.

## 4. Penetration depth in anisotropic samples

Most superconductors of current interest are strongly anisotropic. This presents experimental complications which are important to understand. We consider the simplest case, in which there are now two different London penetration depths, $\lambda_{ab}$ and $\lambda_c$, using the notation relevant to copper oxides. The geometry is shown in Figure 2a. A sample of constant cross section in the x-y plane extends infinitely far in the z-direction. The sample has thickness $2d$ and width $2b$ and length $w \to \infty$. A magnetic field is also applied in the z-direction so demagnetizing corrections are absent. To facilitate comparison with the copper oxides we take the y direction along the c-axis and let the $x$ and $z$ directions correspond to the a and b axes. Choosing the coordinate axes to lie along principal axes of the superfluid density tensor (see 2.) we have

$$\nabla^2 j_i = \frac{j_i}{\lambda_{ii}^2} \qquad (25)$$

so $\lambda_{xx,zz} = \lambda_{ab}$ and $\lambda_{yy} = \lambda_c$. For the case considered here, supercurrents will flow only in the $x$ and $y$ directions. In-plane supercurrents flowing in the $x$ direction penetrate from the top and bottom faces a distance $\lambda_{ab}$. Interplane (*c*-axis) supercurrents flowing in the y direction penetrate from the left and right edges a distance $\lambda_c$.

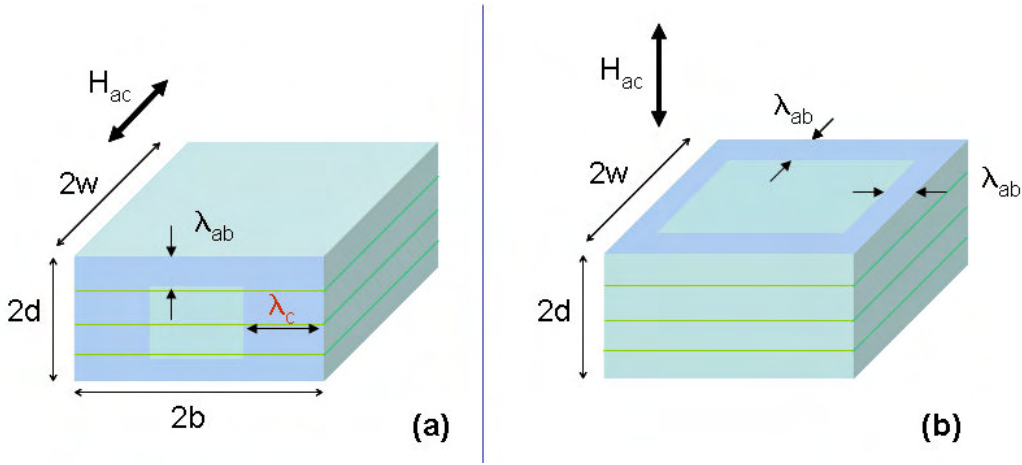

Figure 2. Experimental configurations. (a) geometry relevant to Eq.(26) when $w \to \infty$. (b) finite $w$ with with field normal to conducting planes. In-plane penetration depth is assumed isotropic.

Using (8) and Maxwell's equations it is straightforward to derive the generalized London equation for the magnetic field, [76, 112, 141]

$$\lambda_c^2 \frac{\partial^2 B_z}{\partial x^2} + \lambda_{ab}^2 \frac{\partial^2 B_z}{\partial y^2} = B_z \qquad (26)$$

The equation (26) for the field mixes components of the penetration depth and has a less direct interpretation than Eq. (25). Measurements of the type we will discuss measure a change in the impedance of a resonator and are therefore sensitive to the energy of the sample as a function of its penetration depth. This energy is proportional to an effective magnetic susceptibility, obtained by solving eq.(9) with appropriate boundary conditions. This problem was first solved by Mansky *et al.* [141] who obtain,

$$-4\pi\chi = 1 - \frac{\lambda_{ab}}{d}\tanh\left(\frac{d}{\lambda_{ab}}\right) - 2\lambda_c b^2 \sum_{n=0}^{\infty}\frac{\tanh\left(\tilde{b}_n/\lambda_c\right)}{k_n^2 \tilde{b}_n^3} \qquad (27)$$

where $k_n = \pi(1/2+n)$ and $\tilde{b}_n = b\sqrt{\left(1+\left(k_n \lambda_{ab}/d\right)^2\right)}$. (Note that Eq.(27) has been re-written in a different form compared to [141]). In practice, summation to n = 50 suffices. It is easy to show that Eq.(27) gives the correct result for limiting cases. For example, if $\lambda_{ab} \ll d$ and $\lambda_c \ll b$ we obtain $\tilde{b}_n \approx b$ and with $\sum_{n=0}^{\infty} k_n^{-2} = 1/2$, $-4\pi\chi = 1 - \lambda_{ab}/d \tanh(d/\lambda_{ab}) - \lambda_c/b \tanh(b/\lambda_c)$ as expected. More often one has $\lambda_c \gg \lambda_{ab}$. In that case, for typical crystal dimensions one has $\lambda_c/d \gg \lambda_{ab}/b$ and the susceptibility is dominated by $\lambda_c$. If the interplane transport is incoherent, $\lambda_c$ is considerably more complicated to interpret than the in-plane depth $\lambda_{ab}$ since it involves poorly understood properties of the interplane transport mechanism. [178, 92, 187]

To measure $\lambda_{ab}$ in highly anisotropic materials, one must find either extremely thin samples ( $\lambda_c/d \ll \lambda_{ab}/b$ ) or restrict supercurrents to the conducting planes. The latter approach requires the geometry shown in Figure 2(b), with the magnetic field normal to the planes. In this case, large demagnetization effects occur and no closed form solutions of the London equation exist. Strictly speaking, a demagnetizing factor can only be defined for ellipsoidal samples, but for real samples an effective demagnetizing factor is still a useful concept. A more difficult problem is to determine the effective sample dimension by which to normalize the penetration depth in this demagnetizing geometry. A semi-analytical solution for this problem was found for thin slabs by Prozorov *et al.* [168]. The susceptibility is given by,

$$-4\pi\chi = \frac{1}{1-N}\left[1 - \frac{\lambda}{\tilde{R}}\tanh\left(\frac{\tilde{R}}{\lambda}\right)\right] \qquad (28)$$

where $\tilde{R}$ is effective dimension and $N$ is the demagnetization factor. For a disk of thickness $2d$ and radius $w$ and a magnetic field applied perpendicular to the plane of the disk (i.e. along $d$ ),

$$\tilde{R} \approx \frac{w}{2\left\{1+\left[1+\left(\frac{2d}{w}\right)^2\right]\arctan\left(\frac{w}{2d}\right) - \frac{2d}{w}\right\}} \qquad (29)$$

In the thin limit, $d \ll w$, $\tilde{R} \approx 0.2w$. The demagnetization correction is given by,

$$\frac{1}{1-N} \approx 1 + \frac{w}{2d} \qquad (30)$$

In Eq. (28), the $\tanh(\tilde{R}/\lambda)$ term is only an approximation (it is exact for infinitely long slab of thickness $2w$). For an infinitely long cylinder the London equation gives $I_1(\tilde{R}/\lambda)/I_0(\tilde{R}/\lambda)$ - the ratio of modified Bessel functions of the first kind. However, these distinctions are important only close to $T_c$ (more specifically where $\lambda \geq 0.4w$) and even then the results are quite similar. At low temperatures where $\tilde{R}/\lambda \gg 1$ the hyperbolic tangent factor is essentially unity and therefore irrelevant. For rectangular slabs Eqs.(28), (29) and (30) can be applied with the effective lateral dimension

$$\tilde{w} = \frac{db}{b + 2d/3} \tag{31}$$

Equation (31) was obtained by fitting the numerical solutions of Eq.(27) in its isotropic form ($\lambda_c = \lambda_{ab}$) to Eq.(28). A straightforward generalization of Eq.(29) would lead to a similar expression, but without factor 2/3 in the denominator. Equation (31) is more accurate, because it is obtained by fitting the numerical solutions of Eq. (27) for a rectangular slab, not for the disk described by Eq.(29). We have used the model described by Eqs. (28)-(31) extensively. As this paper will show, the results are generally excellent in terms of signal to noise ratio and when comparison is possible, they agree very well with measurements by other techniques. The necessity of using this measurement geometry has also lead to qualitatively new physics, namely the observation of surface Andreev bound states, as we discuss later. The existence of these states is direct evidence of unconventional superconductivity.

When large enough single crystals are unavailable penetration depth measurements are often done on granular samples. Good results can often be obtained by approximating the grains as spherical and using,

$$-4\pi \chi_{Sphere} = \frac{3}{2}\left(1 - \frac{3\lambda}{r}\coth\left(\frac{r}{\lambda}\right) + \frac{3\lambda^2}{r^2}\right) \tag{32}$$

This is most useful in the limit $\lambda \ll r$. For a more accurate measure Eq. (32) may be averaged over the grain size distribution, $f(r)$, [209, 164]

$$-4\pi \langle \chi \rangle = \frac{\int_0^\infty r^3 f(r) \chi(r) dr}{\int_0^\infty r^3 f(r) dr} \tag{33}$$

If $\lambda$ is anisotropic, as is often the case, the grains can be cast in epoxy and aligned with a large magnetic field prior to measurement. [160, 142]

**5. Absolute value of the penetration depth**

Resonator perturbation methods, whether low or high frequency, provide extremely precise measurements of changes, $\Delta\lambda$, in the penetration depth as a function of an external parameter; typically temperature or magnetic field. The absolute penetration depth is then given by $\lambda(0) + \Delta\lambda(T,H)$. The determination of $\lambda(0)$ itself is still a difficult problem and several methods have been developed.

In principle one could determine $\lambda(0)$ by calculating the expected sample susceptibility as described previously and then measuring the resulting change in frequency as the sample is removed from the resonator. However, Eq. (28) is only approximate and for most situations, $\lambda/\tilde{R} \ll 1$. Moreover the pre-factor cannot be calculated precisely. Therefore, it is not possible to differentiate between a perfectly diamagnetic sample ($\lambda/\tilde{R} = 0$) of arbitrary shape and one with a finite penetration depth if $\lambda/\tilde{R} \ll 1$. For highly anisotropic materials the interplane penetration depth is often large enough, relative to sample dimensions, that it *can* be directly determined by this method. [45, 93] For powder samples in which the grain size distribution is known, Eq. (32) can be used to extract the full $\lambda$ as discussed earlier. [161] For thin films, since the geometry is well controlled, it is also possible to determine the full $\lambda(T)$. This is typically done using a mutual inductance technique with drive and pickup coils on opposite sides of the film, well away from any edges. [66, 125]

It is sometimes possible to obtain $\lambda(0)$ from measurements of the surface impedance $Z_s = R_s + iX_s$. Changes in the frequency and quality factor of a microwave cavity are directly related to $X_s$ and $R_s$ respectively. In the normal state the sample has skin depth δ, DC conductivity $\sigma_{DC}$, and $X_s = R_s = 1/\delta\sigma_{DC}$. In the superconducting state one has $X_s = \omega\mu_0\lambda$. Therefore $X_s(T=0)/R_s(T=T_c^+) = 2\lambda(0)/\delta$. By now measuring the change in $X_s$, upon cooling from above $T_c$ to $T=0$, it is possible to obtain $\lambda(0)$, subject to assumptions about the distribution of microwave currents. [143]

Muon spin rotation (μSR) has been widely used to determine $\lambda(0)$. The details are complicated and we refer the reader to the literature for details. [194] μSR measures a second moment of the magnetic field distribution around a vortex. To extract the penetration depth requires a model for the vortex lattice as well as knowledge of the muon's location. The moment of the field distribution depends in a nontrivial way upon the applied field, so an extrapolation to $H = 0$ is required to yield reliable estimates of $\lambda(0)$. Nonetheless, values of $\lambda(0)$ obtained from μSR do often agree well with those obtained by other methods in cases where a comparison is possible.

In the limit that vortex pinning is negligible, the magnetization of a superconductor in the mixed state is a well defined function of $\lambda$. With suitable corrections for vortex core effects thermodynamic magnetization is given by, [72, 86]

$$M = -\frac{\phi_0}{32\pi^2\lambda^2}\ln\left(\frac{\eta H_{c2}}{H}\right) \quad (34)$$

Experimentally, if one can find a region of field over which *M* is reversible, then this result can be assumed to hold and the penetration depth may be extracted. The difficulty is that only very clean, weakly pinned systems exhibit a sufficiently large interval of reversible behavior and even if they do, it is often non-logarithmic in *H*. Understanding of such deviations involves an analysis of the field dependence of the effective coherence length and possible nonlocal effects. The general analysis of the reversible magnetization applicable to various large-κ superconductors is currently in progress [113].

Infrared reflectivity measurements can also be used to determine $\lambda(T=0)$. This method is useful when anisotropy is an issue, since techniques such as μSR or reversible magnetization average over both directions in the conducting plane. This issue is important in YBCO where the response can be significantly different for currents along the a and b axes. From the reflectance one obtains the conductivity and the frequency dependent penetration depth, $\lambda(\omega) = 4\pi\omega\sigma(\omega)/c^2$. Since the data begins well above $\omega = 0$ one must either extrapolate backward or use a sum rule argument to obtain $\lambda(T=0)$ at $\omega = 0$. [19]

We have developed a differential technique that can be used in conjunction with the tunnel diode oscillator to determine $\lambda(0)$. The sample under study is coated with a lower-$T_c$ material (typically Al) of known thickness and penetration depth.[84, 169] If the film thickness is smaller than its normal state skin depth, then above $T_c(\text{Al})$ the resonator sees only the sample under study. As one cools below the $T_c(\text{Al})$ the film screens the external field from the sample. The change in resonator frequency from $T/T_c(\text{Al}) \ll 1$ to just above $T_c(\text{Al})$ then allows one to extract the penetration depth of the sample.

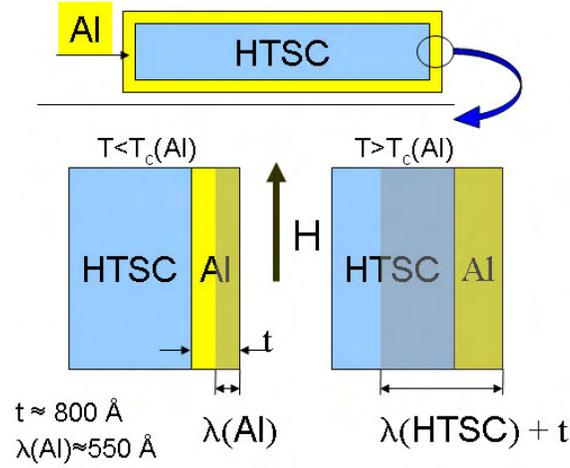

Figure 3. Schematics of the experiment to measure the absolute value of the penetration depth.

Figure 3 explains the experiment. As long as penetration depth of the coated material is sufficiently large, the results will be accurate. The penetration depth is given by,

$$\lambda(\text{HTSC}) = \lambda(\text{Al}) + \frac{\Delta\lambda - t}{1 - \exp(-t/\lambda(\text{Al}))} \quad (35)$$

Figure 4 shows the technique applied to three different superconductors for which the absolute values were also established by other techniques [169]. The results are in a good agreement with the literature values for BSCCO and YBCO and provide a new estimate for the electron-doped superconductor $\text{Pr}_{1.85}Ce_{0.15}CuO_{4-\delta}$ (PCCO). A drawback to the technique is the need to know the penetration depth of the metallic film coating the sample. Although $\lambda(Al)$ is well known for bulk samples, it is certainly possible that in thin films it may differ substantially.

While each method has its virtue, none is completely satisfactory. A reliable, accurate, model-independent technique for measuring $\lambda(0)$ remains an outstanding experimental challenge.

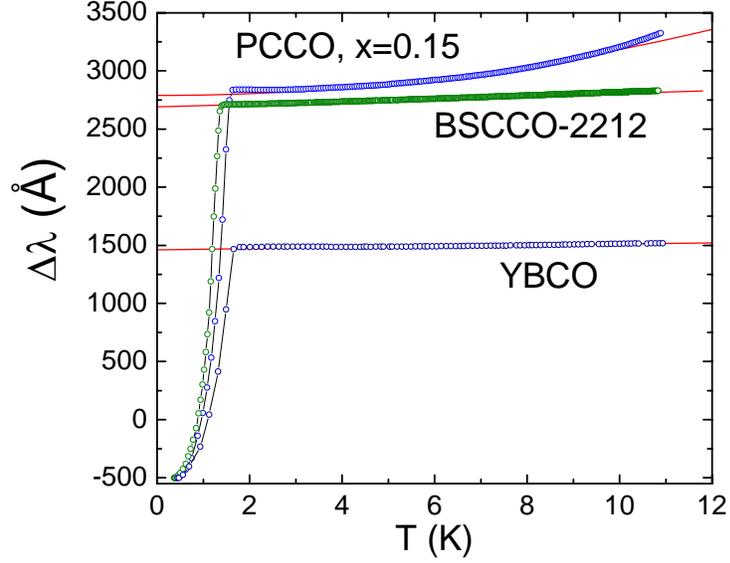

Figure 4. Measurements of the absolute value of the penetration depth in $Pr_{1.85}Ce_{0.15}CuO_{4-y}$ (PCCO), $Bi_2Sr_2CaCu_2O_{8+y}$ (BSCCO-2212) and $YBa_2Cu_3O_{7-y}$ (YBCO). The negative initial values correspond to the thickness of the sputtered aluminium layer. See for details Ref. [169].

## 6. Penetration depth in high-$T_c$ cuprates

6.1. Hole-doped high-$T_c$ cuprates

Early penetration depth measurements in high-$T_c$ cuprates claimed either s-wave pairing, probably due to insufficient sensitivity, or $T^2$ behavior, due to poor quality samples. Microwave measurements on single crystals of $YBa_2CuO_{7-\delta}$ (YBCO) by Hardy *et al.* [87] were the first to show the linear - $T$ dependence characteristic of line nodes. In Figure 5 we show data for a single crystal of optimally doped YBCO, measured with the tunnel diode oscillator at a frequency of 12 MHz. $\rho$ and $\Delta\lambda$ are shown with the ac field parallel to both the a and b crystalline axes. $\rho$ is linear over a substantially wider range than $\lambda$, illustrating the point made earlier that these two quantities have the same T dependence only asymptotically as $T \to 0$. For the optimally doped sample shown here, $d\lambda/dT \approx 4.2$ Å/K in close agreement with the microwave data. [87]

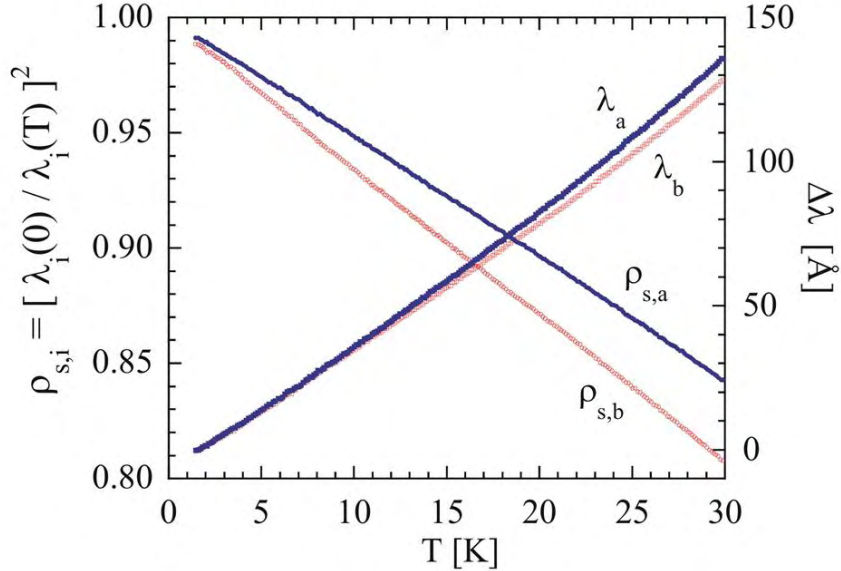

Figure 5. Superfluid density and penetration depth versus temperature in optimally doped YBCO crystal. Data is shown for both a and b directions within the conducting plane. [46]

By now, a linear temperature variation in the superfluid density has also has also been observed in $Bi_2Sr_2CaCu_2O_{8+\delta}$ (Bi-2212) [101, 127], $Tl_2Ba_2CuO_{6+\delta}$ (Tl-2201) [38], $HgBa_2CuO_{4+\delta}$ (Hg-1201), $HgBa_2Ca_2Cu_3O_{8+\delta}$ (Hg-1223) and $La_{2-x}Sr_xCuO_4$ (214) [160, 162]. Fitting (22) to the data on these various materials, including underdoped YBCO, Panagopoulos and Xiang showed that $\Delta_0(0) = 2.14\, k_B T_C$ is approximately obeyed. This is the gap amplitude expected for a weak-coupling d-wave superconductor. The gap function may take more complicated forms consistent with $d_{x^2-y^2}$ symmetry. For example, the form $\Delta = \Delta(\varphi, T)\cos(2\varphi)$ was suggested as a way to reconcile penetration depth data, which is sensitive to the gap amplitude at the nodes, $\Delta(\varphi_{Node}, T)$, with data from probes such as ARPES and tunneling which also probe away from these regions. [160, 162]. Some highly unusual nonmonotonic behavior has been reported for electron-doped cuprates [28, 147] and BSCCO-2201 [116].

6.2. Electron-doped cuprates
While there is very strong evidence for d-wave pairing in the hole-doped materials, the situation in the electron-doped cuprates has been far more controversial, probably owing to the difficulty in growing high quality single crystals. Early microwave data in $Nd_{2-x}Ce_xCuO_{4-y}$ (NCCO) down to 4.2 Kelvin was interpreted within an s-wave model. [6] However, Cooper suggested that the spin paramagnetism of $Nd^{3+}$ ions could be masking a power law temperature dependence expected if the material were d-wave. [55] Lower temperature measurements on single crystals of NCCO clearly showed a vary large spin paramagnetic effect on the penetration depth below 4 K, as we describe later. [170, 2, 171] Additional measurements on nonmagnetic $Pr_{2-x}Ce_xCuO_{4-y}$ (PCCO) down to 0.4 K showed a superfluid density varying as $T^2$. The data are shown in Figure 6, for several crystals. Data for Nb, a fully gapped s-wave superconductor, is shown for comparison.

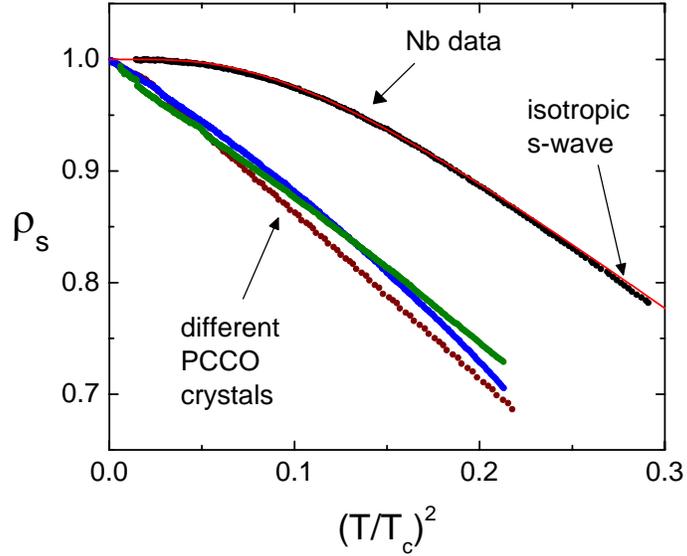

Figure 6. Superfluid density in three PCCO crystals plotted versus $(T/T_c)^2$ to emphasize dirty d-wave behaviour. Data for Nb obtained in the same apparatus is also shown, along with the expected behavior for an isotropic s-wave superconductor.

The quadratic power law is consistent with a d-wave pairing state exhibiting unitary limit impurity scattering, as we discuss later. [91, 170, 114, 171] Coincident measurements of half-integral flux quanta in PCCO films [205] also gave evidence for d-wave pairing in PCCO. Later mutual inductance measurements on PCCO thin films have shown a variety of temperature dependencies ranging from $T^3$ to $T$ and, more recently, exponential, depending upon the method of film growth and existence of a buffer layer. [109] Our own measurements on laser ablated thin films of PCCO continue to show power law behavior that depends upon the oxygen doping level. [193] Figure 7 shows data for optimally doped PCCO film and the fit to the disordered d-wave behavior, Eq.(37), which apparently describes the data very well.

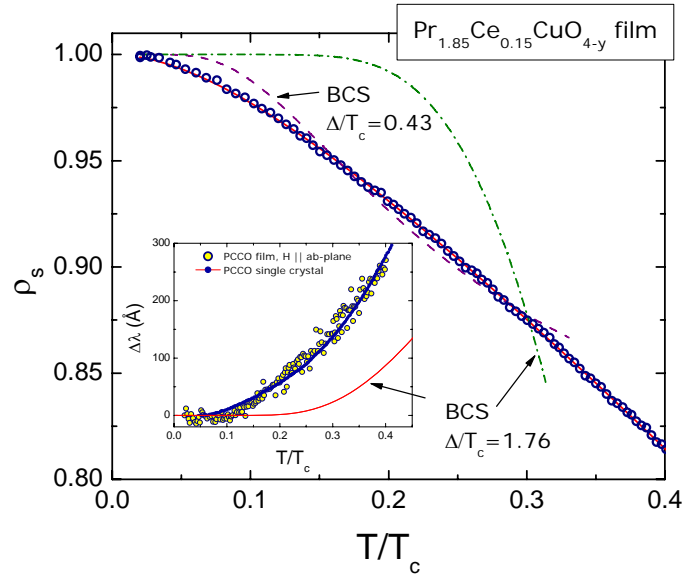

Figure 7 (main frame) Superfluid density in optimally doped PCCO film measured in with $H \parallel c$-axis. s-wave BCS behaviour is shown by dashed lines, while a fit to Eq. (37) is shown by solid line. The inset shows a comparison between single crystal (solid line) and thin film data, the latter measured in the $H \parallel ab$-plane orientation. A standard BCS curve is also shown.

To avoid possible demagnetizing effects in the $H \parallel c$ orientation, we also measured the PCCO films with $H \parallel ab$-plane. For films of order $\lambda$ or less in thickness, the signal is very weak and the data is somewhat noisy, as shown in the inset. Nonetheless, to the best of our knowledge, the inset to Figure 7 is the first reported measurement of a thin film in such orientation. The data is fully consistent with previous measurements on single crystals.

A large number of experimental and theoretical works currently support d-wave pairing in the electron-doped cuprates. Tunneling measurements in PCCO show zero bias conductance peaks. These are now believed to arise from Andreev bound states (discussed later) and are thus direct evidence for unconventional pairing. However, the presence of these states appears to depend upon doping so s-wave pairing for some range of parameters is not ruled out. [26] Raman spectroscopy [177, 133], ARPES [14, 147] specific heat [217] and Hall effect [131] are all consistent with the d-wave picture. A pseudogap, a signature of hole-doped cuprates, has also been observed [147].

Recent works have suggested that the angular dependence of the gap function in e-doped cuprates can differ significantly from the $\Delta(0)\cos(\varphi)$, which appears to describe hole-doped materials. Specifically, the gap is significantly nonlinear in the vicinity of the nodal directions. The maximum gap is not at the Brillouin zone boundary, but at the so-called "hot spot" where the antiferromagnetic fluctuations may strongly couple to the electrons on the Fermi surface, indicating spin fluctuation mediated pairing in electron-doped superconductors. This nonlinearity has been identified experimentally [147, 28] and explored theoretically [216, 105]. The gap angular dependence suggested in Ref. [147] is

$$\Delta(\varphi) = \Delta(0)\left(\beta\cos(2\varphi) + (1-\beta)\cos(6\varphi)\right) \tag{36}$$

with $\Delta(0) = 1.9\,\text{meV}$ and $\beta = 1.43$ determined from direct ARPES measurements [147] for $Pr_{0.89}LaCe_{0.11}CuO_4$ (PLCCO) – an electron-doped superconductor with $T_c = 26\,\text{K}$.

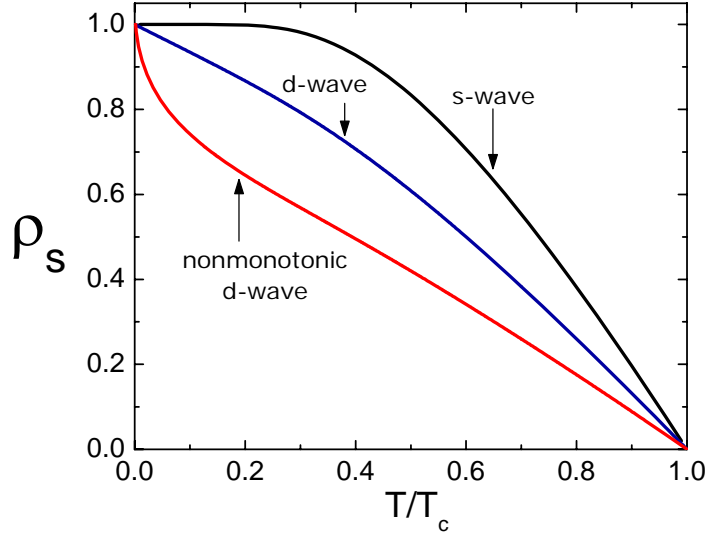

Figure 8. Superfluid density calculated for the non-monotonic gap function, Eq.(36) for $\beta = 1.43$.

We used this function to calculate the gap amplitude using the self-consistent gap equation, see Figure 1 and obtained $\Delta(0)/T_c = 1.19$, which should be compared to the experimental value of 0.85 [147]. These values are not too different from each other and both are much smaller than the weak-coupling s-wave material is 1.76. However, this non-monotonic gap function, Eq.(36) should result in an unusual sub-linear temperature dependence of the superfluid density, shown in Figure 8. On the other hand, it is quite possible that such unusual gap structure comes from the interplay between the superconducting gap and pseudogap, as it was recently observed in BSCCO-2201. [116] In this case, the superfluid density will only be determined by the superconducting gap and penetration depth measurements will be very helpful to elucidate the physics.

## 7. Effect of disorder and impurities

7.1. Non-magnetic impurities
Early measurements of the penetration depth in thin films and some crystals of copper oxide superconductors showed a $T^2$ dependence, instead of the expected linear $T$ dependence for a gap function with line nodes. The problem was resolved by Hirschfeld and Goldenfeld [91, 165] who showed that resonant (unitary-limit) scattering generates a nonzero density of quasiparticle states near $E = 0$. These states lead to a $T^2$ variation of the penetration depth below a crossover temperature $T^*$. The unitary limit was required since Born limit scatterers would lead to a rapid suppression of $T_c$, which was not observed. (See Figure 9 below). A useful interpolation between the linear and quadratic regimes was suggested [91],

$$\lambda(T) = \tilde{\lambda}(0) + \alpha \frac{T^2}{T + T^*} \qquad (37)$$

where $\tilde{\lambda}(0)$ is the effective penetration depth obtained by extrapolation of the linear region of $\lambda(T)$ to $T = 0$. The crossover temperature is given by $k_B T^* \simeq 0.83\sqrt{\Gamma \Delta(0)}$ where $\Gamma = n_i n/\pi N(0)$ is the scattering rate parameter. $n_i$ is the concentration of impurities, n the electron density $n$ and $N(0)$ is the density of states at the Fermi level. $T^* \simeq 0.01 T_c$ is a typical value for high quality YBCO.

Impurities also modify the penetration depth at T = 0 : $\tilde{\lambda}(0)/\lambda(0) \simeq 1 + 0.79\sqrt{\Gamma/\Delta(0)}$. Assuming, for example, a weak coupling BCS result for the d-wave gap, $\Delta(0)/k_B T_c \simeq 2.14$, one has

$$\Gamma \simeq \frac{k_B (T^*)^2}{1.47 T_c} \tag{38}$$

The modification of the zero-temperature penetration depth is then given by,

$$\tilde{\lambda}(0)/\lambda(0) = 1 + 0.95 T^*/\Delta(0) = 1 + 0.44 T^*/T_c \tag{39}$$

which is quite small for clean crystals. This "dirty d-wave" model has been thoroughly studied in both single crystals and thin films of YBCO. [32, 125] In less clean samples the impurity-dominated regime can be a substantial portion of the low temperature region. In that case,

$$\lambda(T) - \tilde{\lambda}(0) = c_2 T^2 \tag{40}$$

where,

$$c_2 = 0.83 \frac{\lambda(0)}{\Delta(0)^{3/2} \Gamma^{1/2}} \tag{41}$$

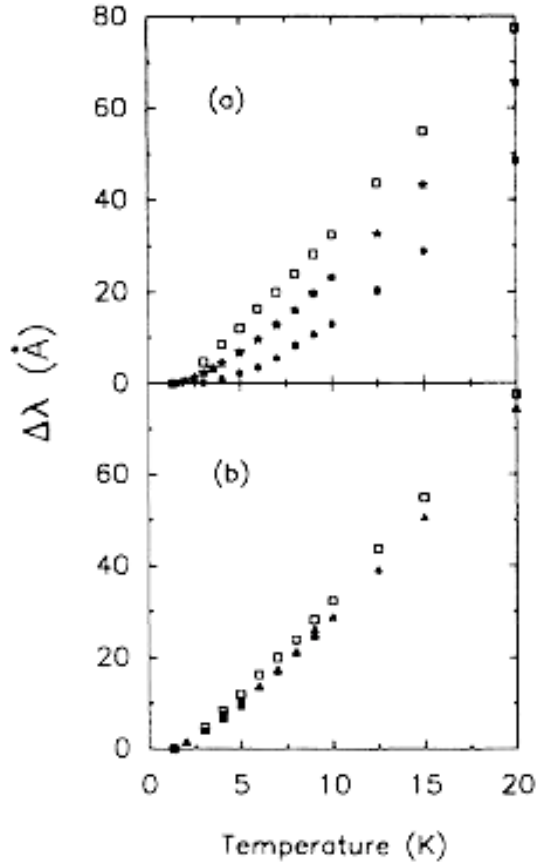

Figure 9. (a) Influence of Zn impurities (resonant scatterers) on the penetration depth for (from top to bottom) Zn%=0, 0.15%, 0.31%. (b) Negligible effect of Ni impurities (Born scatterers) – pure and 0.75%. *Reprinted figure with permission from Ref. [32]. Copyright (1994) by the American Physical Society.*

Figure 9 shows the temperature dependence of the magnetic penetration depth in YBCO single crystals with Zn and Ni substituted for Cu.*[32]*. In the case of Zn, Figure 9 (a), there is a clear evolution from the linear – $T$ behavior to a crossover regime. Ni, shown in Figure 9 (b), does not produce a comparable effect even at twice the Zn concentration. Zn is a unitary-limit scatterrer, whereas Ni is in Born limit, in agreement with the above discussion.

7.2. Magnetic impurities

The ability of magnetic impurities to break pairs leads to profound effects on all superconductive properties. The most familiar is a suppression of the transition temperature, as first calculated by Abrikosov and Gor'kov. Magnetic impurities also affect the penetration depth in a direct way, though a change of permeability when $\mu > 1$. Combining the second London equation, $\mathbf{j} = -\frac{c}{4\pi\lambda^2}\mathbf{B}$, with Maxwell's equations and the constitutive relation $\mathbf{B} = \mu\mathbf{H}$, we obtain a renormalized penetration depth, $\lambda_\mu = \lambda/\sqrt{\mu}$, analogous to the modification of the skin depth in a normal metal. Here $\lambda$ is the London penetration depth without magnetic impurities. $\lambda_\mu$ is the physical length scale over which the field changes. However, the change in resonant frequency of an oscillator or cavity involves a change in energy, which leads to an additional factor of $\mu$. Equation (28) then becomes

$$-4\pi\chi = \frac{1}{1-N}\left[1 - \frac{\mu\lambda_\mu}{\tilde{R}}\tanh\left(\frac{\tilde{R}}{\lambda_\mu}\right)\right] \quad (42)$$

The extra factor of $\mu$ is absent in the argument of the tanh function since the term within the brackets must reduce to $1-\mu$ as $\lambda \to \infty$. At low temperatures the tanh factor becomes unity and the effective penetration depth that one measures is given by

$$\lambda_{eff} = \sqrt{\mu}\lambda \quad (43)$$

This impurity paramagnetism leads to $\mu \sim T^{-1}$ and therefore a minimum in $\lambda_{eff}$. The competing magnetic and superconducting contributions in Eq.(43) played an important role in the determining the pairing state in electron-doped cuprates. Early penetration depth measurements in $Nd_{2-x}Ce_xCuO_{4-y}$ (NCCO) extended to 4.2 K where, coincidentally, the competing temperature dependences in Eq. (43) lead to a minimum in $\lambda_{eff}(T)$. This made the data appear to saturate, as would be expected for s-wave pairing. Cooper was the first to point out that the paramagnetic effect could mask a possible power law dependence for $\lambda(T)$.[55].

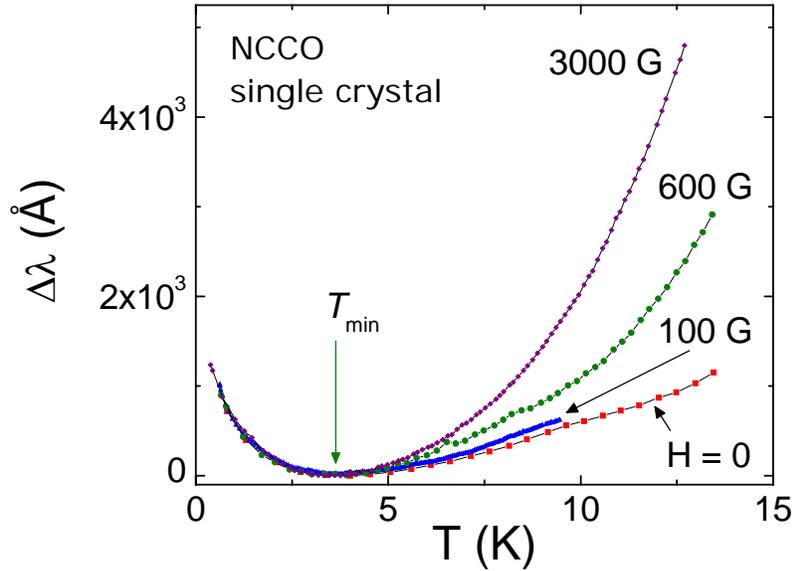

Figure 10. Change in effective penetration depth of electron-doped superconductor NCCO for different applied fields. The change is measured relative to $T_{min}$. For $T < T_{min}$, data at different fields collapse, implying a field-independent permeability for $Nd^{3+}$ ions up to 3 kG.

Figure 10 shows $\Delta\lambda_{eff}(T,B) = \lambda_{eff}(T,B) - \lambda_{eff}(T_{min},B)$ measured in single crystal NCCO at several different magnetic fields aligned parallel to the c-axis. The low-temperature upturn is due to the permeability of $Nd^{3+}$ ions. The fact that the curves collapse onto one plot below $T_{min}$ implies that the permeability is field independent. As the field is increased beyond the values shown here, the spin system will become more polarized and the permeability should decrease. The field independence below $T_{min}$ can help to distinguish an upturn in $\lambda(T_{min})$ from paramagnetic ions from a similar looking upturn due to surface Andreev bound states. As we discuss later, relatively modest fields can quench the upturn from bound states. We have focused on the simplest observable consequence of magnetic impurities on the observed penetration depth. Impurities can have a more profound influence by modifying the gap function itself and thus the entire temperature dependence of the superfluid density. [43]

**8. Organic superconductors**

Probably the most thoroughly studied organic superconductors belong to the class generically referred to as κ-(ET)$_2$X [151, 99, 185, 115, 136] These are highly anisotropic, nearly two dimensional layered layered materials with parameters in the extreme type II limit. NMR measurements show evidence of a spin gap and d-wave pairing. [148, 195], somewhat similar to the situation in the copper oxides. We discuss two of the most widely studied compounds, κ-(ET)$_2$Cu(NCS)$_2$ with $T_c$ = 10.4 K and κ-(ET)$_2$Cu[N(CN)$_2$]Br with $T_c$ = 11.6 K. (The full chemical names are rather lengthy: κ-(ET)$_2$Cu[N(CN)$_2$]Br compound is κ-Di[3,4;3',4'-bis(ethylenedithio)-2,2',5,5'-tetrathiafulvalenium]Bromo(dicyanamido)cuprate(I) and so we use abbreviations.) The chemical structure is no less complex, as shown in the figure.

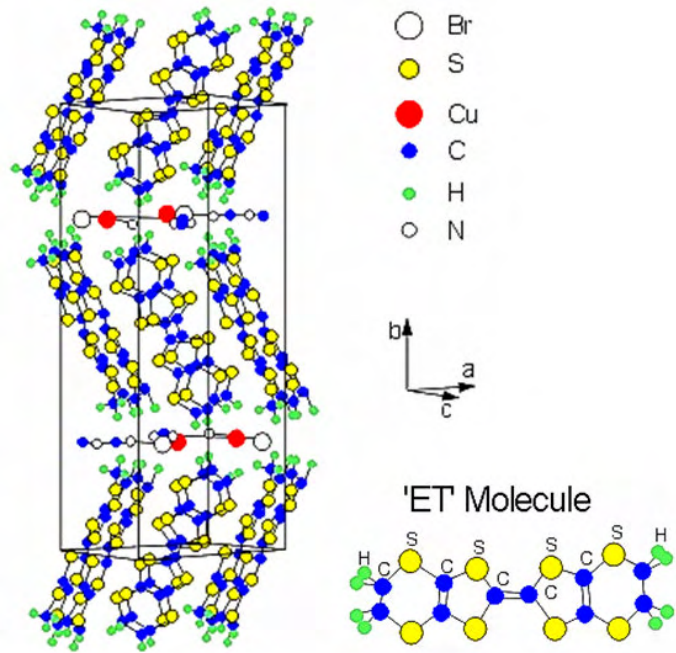

Figure 11. Molecular structure of κ-(ET)$_2$Cu[N(CN)$_2$]Br compound. The building block – the 'ET' molecule is also shown. *Picture – courtesy of John Schlueter (Argonne National Lab)*

Both materials superconduct under ambient pressure. Early penetration depth measurements claimed s-wave pairing, but lower temperature and higher precision measurements by Carrington *et al.* down to 0.35 K provided strong evidence for nodal quasiparticles. [45] The data is shown in Figure 12. The distinction between penetration depth and superfluid density is particularly important here, since $\lambda$ changes rapidly with temperature and its absolute value is large. The measured quantity is $\Delta\lambda = \lambda(T) - \lambda(0.36\,K)$, from which the in-plane superfluid density is calculated for several choices of $\lambda(T \approx 0)$. $\mu SR$ measurements typically give $\lambda(0) \approx 0.8\,\mu m$. [123] For any plausible choice of $\lambda(0)$ the data clearly follows a power law. Fits correspond to Eq. (37) ("dirty d-wave") yield values for the impurity crossover $T^* \approx 0.8\,K$. The fact that the measurements extend down to $T/T_c = 0.03$ rules out all but an extremely small energy gap. By now, a large number of other measurements also indicate d-wave pairing. [13, 100, 166, 21] However, there is still no consensus on the location of nodes on the Fermi surface. And, some specific heat measurements do show evidence for an energy gap in these materials. [64]

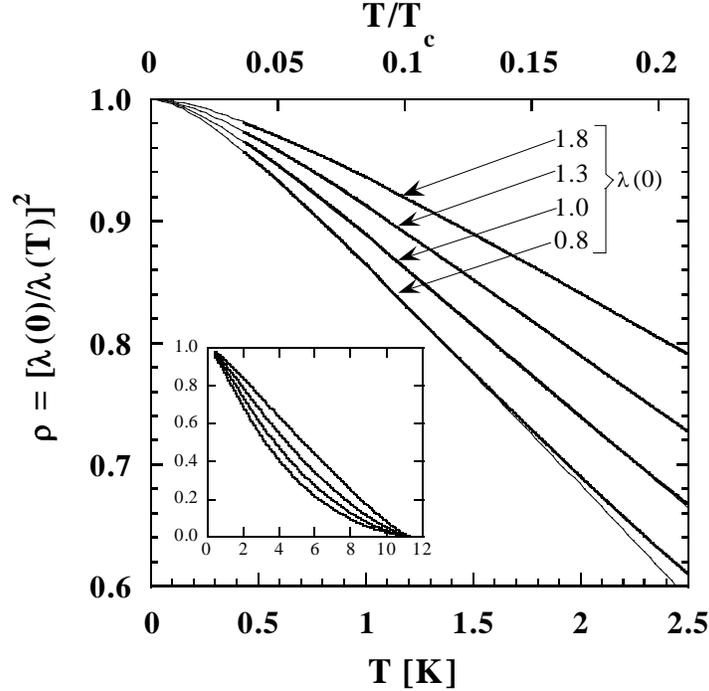

Figure 12. In-plane superfluid density for different choices of $\lambda(0)$ in $\kappa$-(ET)$_2$Cu[N(CN)$_2$]Br. Fits to the data use the dirty d-wave interpolation formula of Eq. (37). Inset shows data up to $T_c$. [45]

Despite this very strong evidence for nodal quasiparticles, the superfluid density is never purely linear and in fact exhibits an interesting regularity. Figure 13 shows the penetration depth itself plotted versus $T^{1.5}$. Over nearly a decade of temperature, samples of both $\kappa$-(ET)$_2$Cu[N(CN)$_2$]Br (a,b) and samples $\kappa$-(ET)$_2$Cu(NCS)$_2$ (c,d) fit this power law with extraordinary precision. Although the dirty d-wave functional form can, with an appropriate choice of parameters, appear as a $T^{1.5}$ power law, the fact that every sample measured obeys this law would imply a remarkable regularity in the impurity crossover temperature $T^*$. We have observed no such regularity in the copper oxides. At the time the data was reported, Kosztin et. al. had proposed a Bose-Einstein/BCS crossover theory for the underdoped copper oxides**.** [49, 118] They predicted a superfluid density that would vary as $T + T^{1.5}$, the new $T^{1.5}$ component coming from finite momentum pairs, similar to a Bose-Einstein condensate. However, our superfluid density data do not fit this power law. As we discuss next, recent measurements on the heavy fermion compound CeCoIn$_5$ also exhibit the same $\Delta\lambda \propto T^{1.5}$ behavior. [159] The authors speculate that the fractional power law behavior may come from a renormalization of parameters near to a quantum critical point. Several different experiments indicate that CeCoIn$_5$ is also a d-wave superconductor, so the experimental situation is somewhat analogous to the organics. .

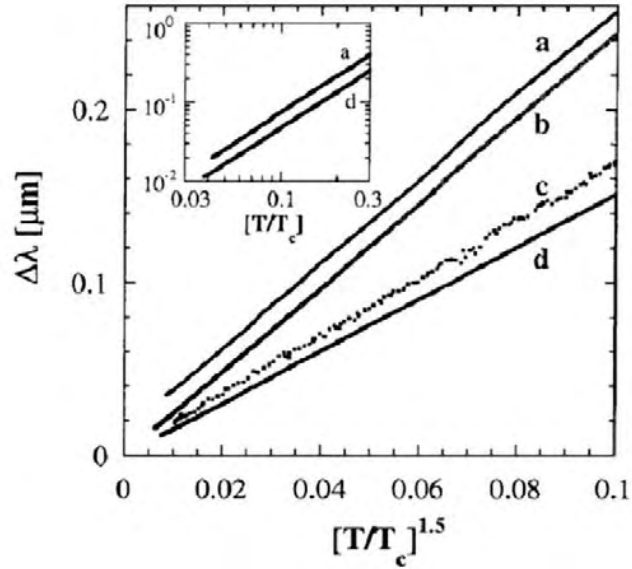

Figure 13. $\Delta\lambda$ for two samples of $\kappa$-(ET)$_2$Cu[N(CN)$_2$]Br (a,b) and two samples $\kappa$-(ET)$_2$Cu(NCS)$_2$ (c,d) plotted versus $T^{3/2}$ [45]

## 9. CeCoIn$_5$

Recently, intense interest has focused on heavy fermion materials of the Ce (M = Co,Ir,Rh)In$_5$ class. Their physics is extremely complex, exhibiting antiferromagnetism, heavy electron behavior, spin fluctuations, highly unusual NMR signals and evidence for d-wave pairing.[57, 83, 90]

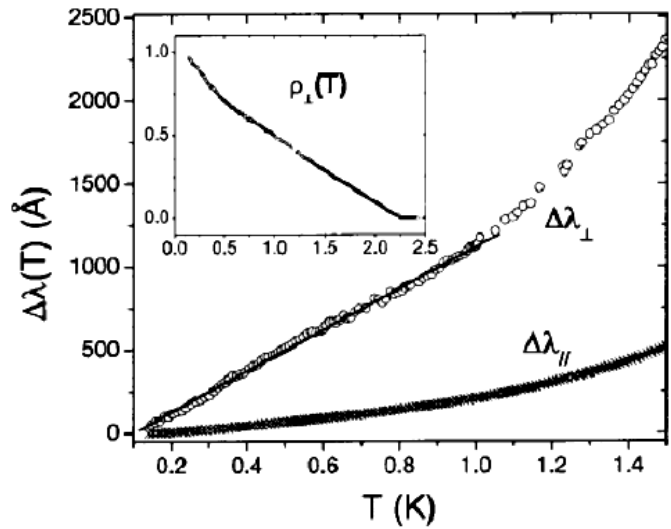

Figure 14. Penetration depth for two orientations in the heavy fermion superconductor, CeCoIn$_5$. Inset shows the superfluid density. *Reprinted figure with permission from Ref. [50]. Copyright (2003) by the American Physical Society.*

Figure 14 shows the penetration depth in CeCoIn$_5$ measured for two different orientations by Chia *et al.* [50]. $\lambda_\perp(T)$ is clearly linear, indicating the existence of line nodes. However, the in-plane penetration depth varies as $\lambda_\|(T) \sim T^{1.5}$. A power law of this type can also be viewed as a combination of $T$ and $T^2$ terms. If the pairing is d-wave, the quadratic term may come from either impurity scattering or nonlocality. Arguing that impurity scattering should contribute to both components while nonlocal corrections apply only to $\lambda_\|(T)$, the authors concluded that the larger power law exponent for the in-plane component is due to the nonlocal response of a superconductor with nodes, as first proposed by Kosztin and Leggett [119]. ( Strong-coupling corrections were also required to make the picture fully consistent.) This same material was also measured by Ozcam et. al., who used a novel micro-coil tunnel diode resonator that probed only the central region of the crystal face. [159] This arrangement eliminates edge effects that could give rise to nonlocal corrections. They also found that $\lambda_\|(T) \sim T^{1.5}$ and concluded that the fractional power law had a more intrinisic origin, possibly related to quantum criticality.[159] We should also mention that CeCoIn$_5$ is the first superconductor where solid evidence exists for the famous high field, inhomogeneous Fulde-Ferrell-Larkin-Ovchinikov (FFLO) state. [121, 70, 23] Tunnel diode oscillator penetration depth measurements have also shown evidence for a transition to this state. [145]

**10. MgCNi$_3$**

The recently discovered non-oxide perovskite superconductor MgCNi$_3$ [89] ($T_c$ = 7.3 K) is viewed as a bridge between high-$T_c$ cuprates and conventional intermetallic superconductors. This material is close to a magnetic instability on hole doping. It was suggested that strong magnetic fluctuations may lead to unconventional pairing [180]. The current experimental situation is controversial. On the one hand, evidence for conventional s-wave behavior is found in specific heat measurements [132], although the authors disagree on the coupling strength. The $^{13}$C nuclear spin-lattice relaxation rate ($1/T_1$) also exhibits behavior characteristic of an s-wave superconductor [190]. On the other hand, a zero-bias conductance peak has been observed, which can be attributed to Andreev bound states and not intergranular coupling or other extrinsic effects [144]. Nonmagnetic disorder introduced by irradiation was found to significantly suppress superconductivity [104]. Such a suppression is not expected in materials with a fully developed gap, so there is a strong indication of an order parameter with nodes. Theoretical calculations support this conclusion [82]. Furthermore, recent theoretical developments predict the possibility of a unique unconventional state [208], which might reconcile the apparently contradictory experimental observations. Precise measurements of the London penetration depth are therefore very important.

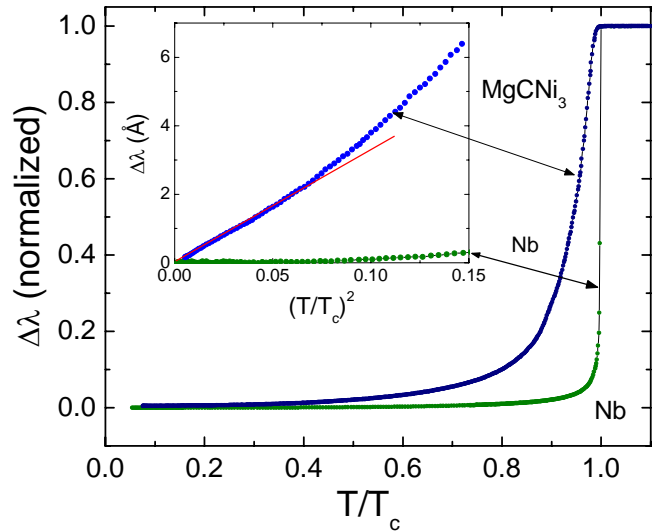

Figure 15. Penetration depth measured in MgCNi$_3$ superconductor. Main frame shows a conventional superconductor (Nb) for comparison. Inset shows the low temperature regime and quadratic behaviour of the penetration depth.[175]

Figure 15 shows measurements of the magnetic penetration depth in MgCNi3 superconducting powder in paraffin. An inset shows low-temperature behavior plotted versus $\left(T/T_C\right)^2$. Such behavior requires a superconducting gap with nodes. Detailed data analysis allowed us to eliminate the possibility of spurious effects such as grain size or transition temperature distribution. [175]

## 11. Spin Triplet pairing

The extension of BCS theory to spin triplet pairing was explored in several pioneering papers written during the 1960's. [15, 5, 200] Pairing states discussed by those authors turned out to precisely describe the stable phases of superfluid $^3$He. [158] Superfluid $^3$He provides undoubtedly the most thoroughly studied and unequivocal example spin triplet, p-wave pairing found in nature. [128] The experimental situation for p-wave superconductors is far more controversial. Candidate materials for p-wave pairing typically have low ($\sim 1K$) transition temperatures so that measurements must contend with self-heating and difficult thermometry issues. Sample purity and surface quality have also been problematic.

An analysis of the allowed p-wave states for a given crystal symmetry can become a complex exercise in group theory.[8] We restrict ourselves to a few simple examples. The pairing state is categorized by the vector $\mathbf{d}(\mathbf{k})$ (see Eq. (2)) which can be considered the spin quantization axis. The quasiparticle energy is given by $E_k^2 = \varepsilon_k^2 + |\mathbf{d}(\mathbf{k})|^2 \pm |\mathbf{d}(\mathbf{k})^* \times \mathbf{d}(\mathbf{k})|^2$ where $\varepsilon_k$ is the usual band energy (measured relative to the Fermi energy) and the sign depends upon the spin of the quasiparticle. The Balian-Werthammer (BW) state that describes the B-phase of superfluid $^3$He has $\mathbf{d}(\mathbf{k}) = \Delta_0 \mathbf{k}$, which has a finite energy gap $\Delta_0$ everywhere on the Fermi surface. In such a state, the penetration depth would vary exponentially at low temperatures, just as in a conventional superconductor. The polar state is defined by $\mathbf{d}(\mathbf{k}) = \mathbf{d}_0 \left(\mathbf{k} \cdot \hat{\mathbf{l}}\right)$ where $\hat{\mathbf{l}}$ is the axis of symmetry for the gap. This state has a line of nodes in the equatorial plane perpendicular to $\hat{\mathbf{l}}$. The energy gap for quasiparticles is given by $\Delta(\mathbf{k},T) = \Delta_0(T) |\mathbf{k} \cdot \mathbf{l}|$. The axial state is defined by $\mathbf{d}(\mathbf{k}) = \mathbf{d}_0 \mathbf{k} \cdot (\hat{\mathbf{e}}_1 + i\hat{\mathbf{e}}_2)$ where $\hat{\mathbf{e}}_1, \hat{\mathbf{e}}_2$ are unit vectors and $\hat{\mathbf{l}} = \hat{\mathbf{e}}_1 \times \hat{\mathbf{e}}_2$. This state describes the A-phase of superfluid $^3$He where $\hat{\mathbf{l}}$ is somewhat like a pair angular momentum vector. For the axial state, the energy gap can be written as $\Delta(\mathbf{k},T) = \Delta_0(T)|\mathbf{k} \times \hat{\mathbf{l}}|$. For a 3D Fermi surface there are two point nodes, but for a 2D Fermi surface this gap is finite everywhere.

In general, the situation is complicated because the supercurrent is no longer parallel to the vector potential. The electromagnetic response now depends on the mutual orientation of the vector potential $\mathbf{A}$ and $\hat{\mathbf{l}}$, which itself may be oriented by surfaces, fields and superflow. A detailed experimental and theoretical study for the axial and polar states was presented in Ref. [63, 85]. The following low-temperature asymptotics were obtained for different orientations of $\mathbf{A}$ and $\hat{\mathbf{l}}$:

$$\rho_{\|,\perp} = 1 - a_{\|,\perp} \left(\frac{T}{\Delta(0)}\right)^{n_{\|,\perp}} \tag{44}$$

The exponents for different orientations are given in Table 1.

Table 1: various low-temperature coefficients for the p-wave pairing state in 3 dimensions

|  | orientation | a | n |
|---|---|---|---|
| Axial: $|\mathbf{k} \times \hat{\mathbf{l}}|$ (two point nodes) | $\hat{\mathbf{l}} \| \mathbf{A}$ | $\pi^2$ | 2 |
|  | $\hat{\mathbf{l}} \perp \mathbf{A}$ | $\dfrac{7\pi^4}{15}$ | 4 |
| Polar: $\mathbf{k} \cdot \hat{\mathbf{l}}$ (equatorial line node) | $\hat{\mathbf{l}} \| \mathbf{A}$ | $\dfrac{27\pi\varsigma(3)}{4}$ | 3 |
|  | $\hat{\mathbf{l}} \perp \mathbf{A}$ | $\dfrac{3\pi \ln 2}{2}$ | 1 |

Heavy fermion superconductors have long been considered possible candidates for p-wave pairing. [71] In 1986, Einzel *et al.* and Gross *et al.* reported measurements of $\lambda(T)$ in UBe$_{13}$ ($T_c = 0.86 K$) down to 60 mK [63, 85] Their experiment utilized a novel SQUID bridge circuit to record the change in sample magnetic moment in a small dc field as the temperature was varied. To our knowledge, this experiment was the first penetration depth measurement to report evidence of unconventional superconductivity in any material. They found $\lambda(T) - \lambda(0) \sim T^2$, as shown in Figure 16. These data are consisent with an axial state. [85, 48, 84, 189].

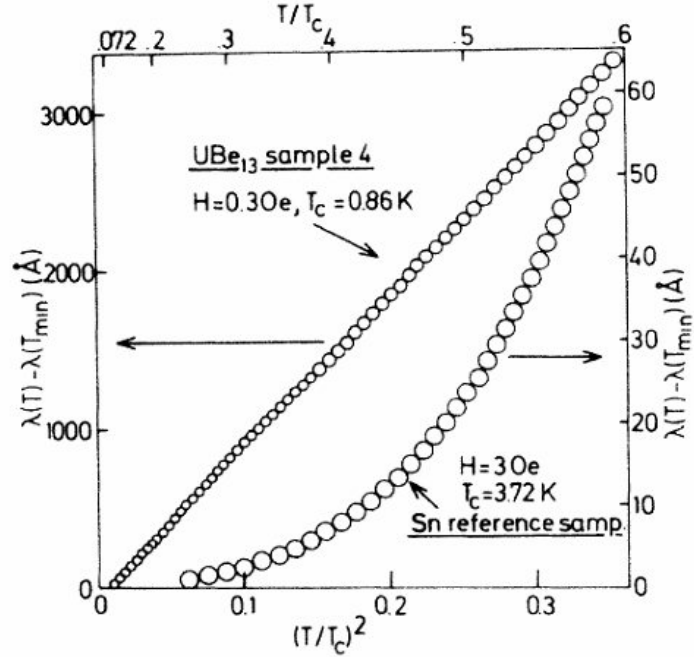

Figure 16. Change in the penetration depth in heavy fermion superconductor UBe$_{13}$. *Reprinted figure with permission from Ref. [63]. Copyright (1986) by the American Physical Society.*

The superconductor $UPt_3$ is one of the strongest candidates for triplet pairing among the heavy fermion materials. [103] There have been several penetration depth penetration depth measurements on this system, all of which show power law behaviour, See Ref. [48, 84, 189] and references therein. The observed power laws range from $T$ to $T^4$. However, the power law was found to vary depending upon the measurement technique, sample surface preparation and frequency. It is fair to say that penetration depth measurements in this material show strong evidence for nodes but are inconclusive beyond that. Theories for the pairing state in $UPt_3$ are complex, involving mixed order parameters and we refer the reader to the literature for fuller discussions. [183, 103]

The possibility of spin triplet, p-wave pairing in SrRuO$_4$ ( $T_c=1.4 K$ ) has generated considerable interest. [139, 137] SrRuO$_4$ is a perovskite with strong anisotropy and strong ferromagnetic fluctuations. The latter are expected to favor triplet pairing.[179, 18]**;** Measurements show a Knight shift that remains constant down to $T/T_c \ll 1$, as expected for a spin triplet state for the field orientation used. [98] Bonalde e*t al.* measured the penetration depth down to 40 mK using a tunnel diode oscillator.[31] Several samples were measured. At the lowest temperatures the superfluid density varied quadratically with temperature, suggesting the presence of nodes. However, other experiments in SrRuO$_4$ point to a state with a finite energy gap. [137] One widely considered candidate is the axial state $\vec{d}(\mathbf{k}) = \Delta_0 \hat{z}(k_x + i k_y)$ where **z** is perpendicular to the basal plane. The authors ruled out "gapless" behavior from impurities which can give a $T^2$ dependence even to a superconductor with an energy gap.[192] They concluded that the quadratic dependence may arise from nonlocality. The issue is not settled and several other triplet pairing states have also been investigated for this material [10, Won, 2000 #217].

Finally, we mention that there is evidence from upper critical field measurements that the one-dimensional organic superconductor (TMTSF)$_2$PF$_6$ has triplet pairing. [124] Experiments on this material require high pressure and low temperatures $T_c=1.13 K$. We are not aware of penetration measurements on this material.

## 12. New features specific to unconventional superconductivity

12.1. Nonlinear Meissner effect

In the presence of a superfluid velocity field $\mathbf{v}_s$ the energy of a Bogolubov quasiparticle is changed by $\delta E_{QP} = \mathbf{v}_s \cdot \mathbf{p}_F$ where $\mathbf{p}_F$ is the Fermi momentum. This effect is sometimes called the quasiparticle Doppler shift. Quasiparticles co-moving with $\mathbf{v}_s$ are shifted up in energy while those moving counter to $\mathbf{v}_s$ are shifted down. For $T > 0$ the increased population of counter-moving quasiparticles constitutes a paramagnetic current that reduces the Meissner screening. The temperature dependence of $\lambda$ ultimately derives from this fact. As $\mathbf{v}_s$ is increased two things occur. First, higher order corrections to the thermal population difference become more important and second, pair breaking effects reduce the gap itself. In a superconductor with a finite energy gap everywhere on the Fermi surface, the supercurrent acquires a correction term quadratic in $\mathbf{v}_s$, $\mathbf{j}_S = -e\rho \mathbf{v}_s \left(1 - \alpha(T)(v_s/v_c)^2\right)$ where $v_c$ is the bulk critical velocity. Since $\mathbf{v}_s$ is proportional to the applied magnetic field $H$, this nonlinearity results in a field-dependent penetration depth,

$$\frac{1}{\lambda(T,H)} = \frac{1}{\lambda(T)}\left[1 - \frac{3\alpha(T)}{4}\left(\frac{H}{H_0(T)}\right)^2\right] \tag{45}$$

$H_0(T)$ is of the order of thermodynamic critical field. At low temperatures the coefficient $\alpha(T) \sim \exp(-\Delta(0)/T)$. [213] This dependence occurs because the Doppler shift must contend with a finite energy gap and so does not affect the quasiparticle population at $T = 0$. The field dependent correction is extremely small since the penetration depth itself is already exponentially suppressed. Any attempt to observe this effect in a conventional type I superconductor must also take account of the very large field dependence that occurs in the intermediate state.

In 1992, Yip and Sauls [215] showed theoretically that the situation would be quite different in a d-wave superconductor. The existence of nodes in the gap function implies that the Doppler shift can change the quasiparticle population at arbitrarily low temperatures so long as $|\delta E_{QP}| \gg k_B T$. In fact, the effect was predicted to be *strongest* at $T = 0$ and to depend upon the orientation of $\vec{v}_s$ relative to the nodal directions. For a d-wave state at $T = 0$, the nonlinearity leads to a non-analytic correction to the current-velocity relation, $\mathbf{j}_S = -e\rho \mathbf{v}_s \left(1 - |\mathbf{v}_s|/v_0\right)$ where $v_0$ is of order the bulk critical velocity. This correction leads to a linear increase in the penetration depth as a function of field. For a d-wave pairing state at $T = 0$ the result is, [215, 213]

$$\frac{1}{\lambda(T=0,H)} = \frac{1}{\lambda(0)}\left[1 - \frac{2}{3}\frac{H}{H_0}\right] \qquad \vec{H} \parallel node$$

$$\frac{1}{\lambda(T=0,H)} = \frac{1}{\lambda(0)}\left[1 - \frac{1}{\sqrt{2}}\frac{2}{3}\frac{H}{H_0}\right] \qquad \vec{H} \parallel antinode \tag{45}$$

$H_0 = 3\phi_0/\pi^2 \lambda \xi$ is of the order of the thermodynamic critical field. The great appeal of this idea lies in the possibility of verifying both the existence of nodes and locating them on the Fermi surface. Yip and Sauls coined the term "nonlinear Meissner effect" (NLME) to describe the phenomenon. As conceived, it results from the field-induced change in quasi-particle populations and does not include field-induced pairing breaking effects on the gap itself. Since the NLME depends upon the quasiparticle energy, and therefore $|\Delta(\mathbf{k})|^2$, it is a probe of nodes but it is not inherently sensitive to the phase of the order parameter.

Despite considerable experimental efforts, the NLME has proven extremely difficult to identify. A large number of constraints must be satisified. First, H must be smaller than the lower critical field to avoid contributions from vortex motion which can also give a linear correction to $\lambda$. Using YBCO as an example, $H_{c1}/H_0 \sim 10^{-2}$. With that restriction, the maximum change in $\lambda$ is predicted to be of order 1% of $\lambda(0)$. Second, unitary limit impurity scattering is predicted to rapidly destroy the NLME so temperatures *above* the impurity crossover $T^*$ are needed. For the best YBCO, $T^* \approx 1$ K. Third, the field dependence is maximal at T = 0 and decreases rapidly once $\delta E_{QP} << k_B T$. For YBCO, this inequality restricts observation of the effect to temperatures *below* 3- 4 K, even at $H = H_{C1}$, the maximum possible field. For higher temperatures the field dependence becomes quadratic and small. The decrease of the linear field dependence with temperature is, however, a distinguishing feature of the NLME for a d-wave state. This point has been ignored in several attempts to identify the effect. These conditions place extremely tight constraints on the observability of the NLME. Several different experiments have been undertaken. The first type focuses on the predicted anisotropy in the penetration depth and therefore the magnetic moment of a crystal. Bhattacharya et.al.[22, 219], rotated a sample of YBCO and searched for harmonics in the angular dependence of the signal indicative of the nodal anisotropy. They observed anisotropy but it was well below the predicted amount.

The second class of experiments directly measure the penetration depth in a dc magnetic field superimposed on a much smaller ac measurement field. Penetration depth measurements by Maeda et. al. [138] first reported a linear field dependence but did not address the question of the temperature dependence. Measurements by Carrington et. al. and Bidnosti et. al.[46, 25] are shown below. In each case, the total change in $\lambda$ was comparable to that predicted for the NLME. Depending upon the sample, field dependences ranged from linear to quadratic and in some cases close to $H^{1/2}$ (right panel). Bidnosti et. al. found the dependence upon sample orientation to be at odds with theory. Neither experiment reported the temperature dependence expected for the NLME. The left panel shows almost no dependence on temperature. The right panel shows a field dependence that *increases* with temperature. Vortex motion (discussed later) can easily lead to either an H or $H^{1/2}$ field dependence that increases with temperature as vortices become more weakly pinned. [46]. The fact that these extrinsic effects are of similar magnitude to the predicted NLME make its observation particularly difficult.

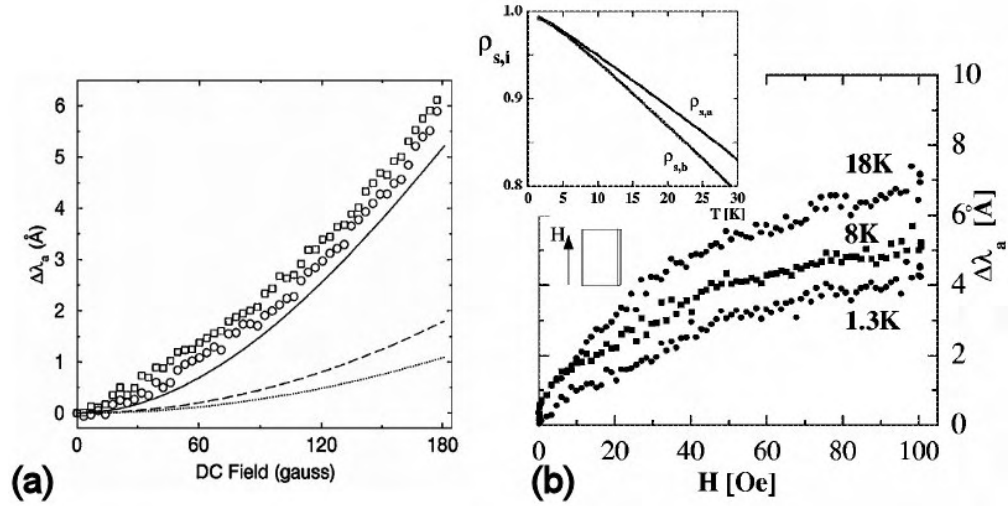

Figure 17. Attempts to identify nonlinear meissner effect in YBCO crystals. **(a)** circles and squared show data for T=4.2 and 7 K, respectively. Solid line, dashed line and dot-dashed line is theoretical expectation for 1.2, 4.2 and 7 K, respectively. *Reprinted figure with permission from Ref. [25]. Copyright (1999) by the American Physical Society.* **(b)** similar measurements from Ref. [46]. Note the scale on the y-axis – both graphs represent state of the art measurements of the penetration depth and both failed to find the expected behaviour.

The inability to observe the NLME lead to a re-examination of the original Yip-Sauls argument and to other suggestions for detecting nonlinear effects. Li *et al.* showed that if the vector potential varies spatially with wave-vector $\vec{q}$ then nonlocal effects suppress the NLME whenever $\vec{v}_s \cdot \vec{q} \neq 0$. This situation occurs when the field is oriented normal to the conducting planes. The suppression occurs for fields $H < H_{C1}$ which effectively renders the effect unobservable. [130] However, for other field orientations nonlocal effects should not be present. Experiments have been done in several orientations and to our knowledge, this orientation dependence has not been identified. The tunnel diode method used by the authors was originally developed to search for the NLME. We have routinely searched for a NLME in several different hole and electron-doped copper oxides and in the organic superconductors discussed previously. All of these materials show clear evidence, through $\lambda(T)$, for nodal quasiparticles. All show a linear variation, $\Delta\lambda(T,H) \sim \beta(T)H$ with field. However, in all cases, $\beta(T)$ varies with temperature in a manner expected for vortex motion, despite applied fields as low as a few Oe.

Dahm *et al.* proposed to exploit the *analytic* corrections to the supercurrent that vary as $v_s^2$ in order to identify the d-wave state. [59] These terms lead to changes in $\lambda$ that vary as $H^2/T$, are apparently less affected by impurity scattering and can be observed over a wider temperature range. As with the linear-in-$H$ [215]) non-analytic corrections, the quadratic corrections are largest at $T=0$ and are thus distinguishable from nonlinear effects in an s-wave superconductor. The nonlinear penetration depth leads to harmonic generation and intermodulation frequency generation and so may have relevance in microwave and mixer applications.

Recently, intermodulation measurements were performed on a number of microwave stripline resonators made from YBCO films. As shown in Figure 18, some films exhibited the $\lambda \sim T^{-1}$ upturn predicted for the nonlinear penetration depth in a d-wave state. [157]

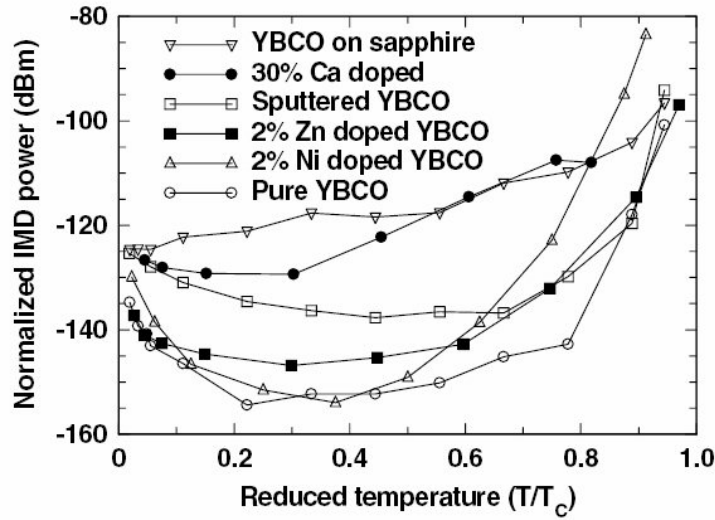

Figure 18. Intermodulation distortion (proportional to $\lambda^2$) for several YBCO films showing the upturn predicted for a d-wave state. *Reprinted figure with permission from Ref. [157]. Copyright (2004) by the American Physical Society*

Figure 18 shows the normalized intermodulation power for several YBCO thin films. There is an obvious non-monotonic temperature dependence. These experiments are the first to observe the low temperature increase in nonlinearity expected for a d-wave superconductor. However, the effect is distinct from the non-analytic, linear-in-*H* behavior first predicted by Yip and Sauls. To our knowledge, the latter has not yet been convincingly identified..

12.2. Surface Andreev bound states

The formulas given previously for superfluid density involve only the absolute square of the gap function. It was therefore believed for some time that penetration depth measurements were insensitive to the phase of the gap function. It was shown theoretically by [16] and experimentally [176] and [47] that penetration depth experiments can in fact be phase sensitive. Unconventional superconductors support the existence of zero energy, current carrying surface Andreev bound states (ABS). [39] These states are a direct consequence of the sign change in the d-wave order parameter. [95]**.** The zero-bias conductance peak widely observed in ab-planar tunneling is generally associated with Andreev bound states states. [12] As shown in Figure 19, for $d_{x^2-y^2}$ symmetry, the effect is maximal for a (110) orientation where the nodal directions are perpendicular to the sample surface. A quasiparticle travelling along the trajectory shown by the arrows initially feels a negative pair potential. After reflection it travels in a positive pair potential. This process leads to zero energy states that are localized within a few coherence lengths of the surface and carry current parallel to it. As such, they can affect the Meissner screening. The effect is absent for the more common (100) orientation.

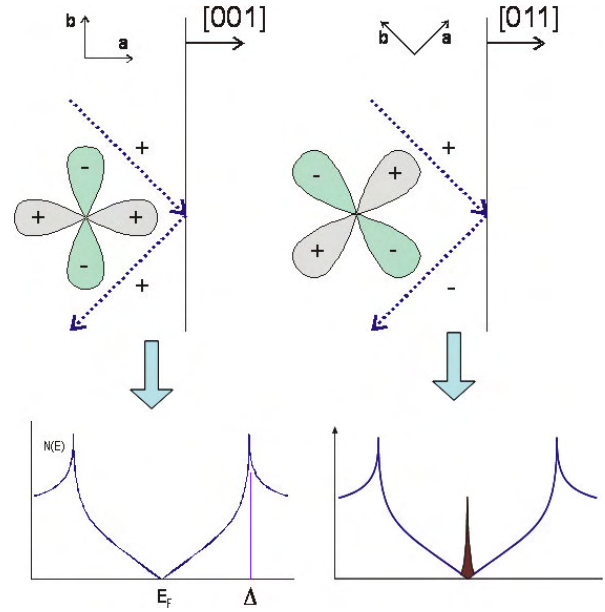

Figure 19. Schematic of the origin of surface Andreev bound states. Magnetic field is normal to the surface. Quasiparticle travelling along the trajectory shown by the arrows experiences a sign change in the pair potential **(right)**, resulting a zero energy bound state localized near the surface. In the other situation **(left)** the effect does not exist. Lower panel shows corresponding response of the density of states – proliferation of the zero-energy bound (Andreev) states.

Andreev bound states may be observed in penetration depth measurements by orienting the magnetic field along the c-axis, inducing shielding currents that flow along the (110) edges of the sample. This geometry was shown in Fig. (2b). Bound states contribute a singular piece to the overall density of current carrying states, $N_{ABS}(E) \sim \delta(E)$. When inserted into Eq.(7), this results in a paramagnetic contribution to the penetration depth, $\Delta\lambda_{ABS} \sim 1/T$. This divergent term competes with the linear T dependence from nodal quasiparticles, leading to a minimum in the penetration depth at $T_m \sim T_c\sqrt{\xi_0/\lambda_0}$. For YBCO $T_m \sim 10\,K$ for a sample all of whose edges have (110) orientation. The effect is shown in Figure 20 where four YBCO crystals with differing amounts of [110] surface were measured. In each case the AC magnetic field was first oriented parallel to the conducting plane (denoted $\lambda_{ab}^{a,b}$) and then along the c-axis (denoted $\lambda_{ab}^{c}$). The first shows the familiar linear variation characteristic of nodal quasiparticles. The second shows the 1/T upturn from bound states. The bound state signal is largest for the sample exhibiting a largest amount of [110] surface.

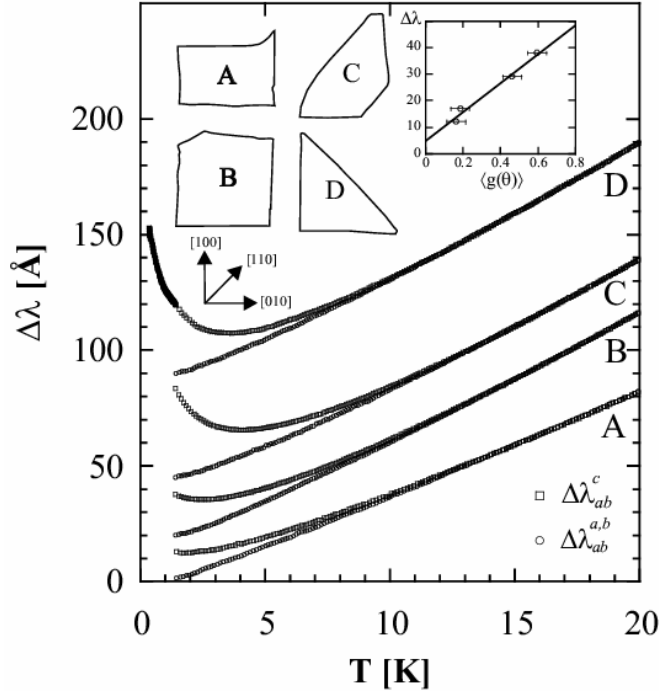

Figure 20 $\lambda_{ab}$ in 4 YBCO crystals. Traces with AC field along c-axis show 1/T upturn while traces taken with AC field along conducting planes show no upturn. Inset shows the relative size of 1/T upturn versus amount of [110] perimeter surface. $<g(\theta)>$. Low temperature portion of trace D was taken in a different apparatus from the portion above 1.3 K. $\lambda_{ab}^c$ denotes in-plane penetration depth with ac field along the c-axis.

Impurity scattering broadens the zero energy peak and reduces the paramagnetic upturn. Another striking feature of the bound state signal is its rapid disappearance with an applied DC magnetic field. Crudely speaking, the field Doppler shifts the zero-energy states, $N(E) \sim \delta\left(E - \frac{e}{c}\mathbf{v}_F \mathbf{A}\right)$, resulting in a field-dependent penetration depth, (see Eq. (7)),

$$\Delta\lambda_{ABS}(T,H) \sim \frac{1}{T\cosh^2\left(H/\widetilde{H}(T)\right)} \tag{46}$$

where $\widetilde{H} \simeq H_c T/T_c$ and $H_c$ is the thermodynamic critical field. Figure 21 shows the field dependence of the 1/T upturn. Fits to both the single quasiparticle trajectory model (Eq. (46)) and the full model of Barash et. al. [16] are also shown. The agreement is remarkably good.

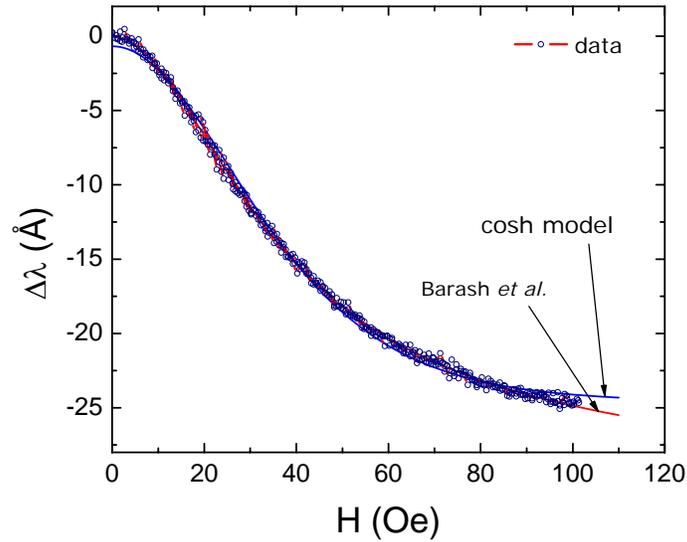

Figure 21. Paramagnetic upturn, $\Delta\lambda(H)$, measured at T = 1.34 K in YBCO crystal. "cosh " refers to Eq. (46) while "Barash *et al.*" is a fit to Ref. [16], Eq.(26).

These highly distinctive temperature, orientation and field dependences differentiate the signal due to bound states from the paramagnetic upturn due to magnetic impurities discussed earlier. Andreev bound states have also been observed in other superconductors. Figure 22 shows the effect in single crystal Bi-2212 for several values of the magnetic field, showing the quenching effect just described.

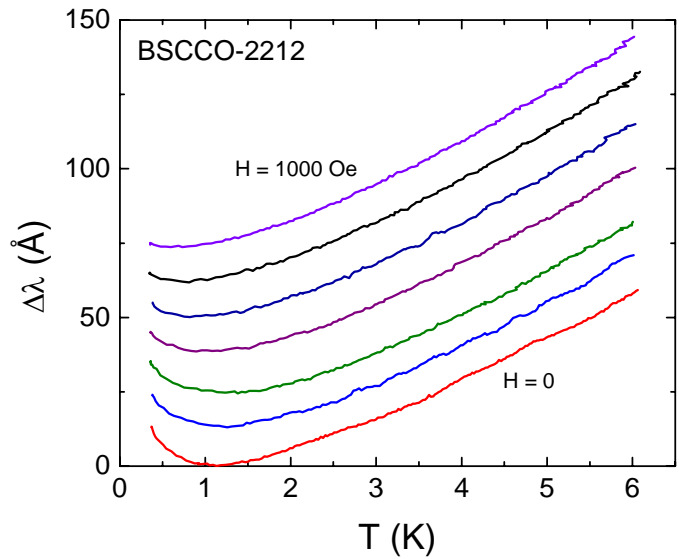

Figure 22. Andreev bound states in Bi-2212 single crystal. The curves are offset for clarity.

12.3. Nonlocal electrodynamics of nodal superconductors

Our discussion to this point has assumed electrodynamics in the London limit for which the **j** is proportional to the local vector potential **A**. This limit holds so long as $\lambda_L \gg \xi = \hbar v_F/\pi\Delta_0$ where $\xi$ is the BCS coherence length. [203] In YBCO for example, one has $\lambda_L(0)/\xi_0 \approx 100$ so the London limit is easily satisfied over the entire temperature range. The issue of nonlocality arose early in the history of superconductivity. Experiments on pure metals gave values of the penetration depth larger than predicted by the London formula, implying that some electrons remained normal. Pippard [163] was the first to suggest a non-local version of London electrodynamics similar to earlier generalizations of the Ohm's law. In this case one has $j_i(\mathbf{x}) = \int K_{ij}(\mathbf{x},\mathbf{y}) A_j(\mathbf{y}) d\mathbf{y}$ where $K$ is the response kernel. If the coherence length (first introduced by Pippard) is larger than the London penetration depth, the response of the superconductor to a magnetic field is weakened due to reduction of the vector potential over length $\lambda_L$. The effective magnetic penetration depth increases [203]

$$\lambda_{eff} \approx a\lambda_L \left(\frac{\xi_0}{\lambda_L}\right)^{1/3} \tag{47}$$

Here $a = 0.65$ in an extreme the type-I situation of $\lambda_L(0) \ll \xi_0$. Since all non-elemental superconductors are type-II, nonlocality would not seem to be an issue for these materials.

Kosztin and Leggett [119] first pointed out that since the BCS coherence length defined as $\xi(\mathbf{k}) = \hbar v_F/\pi\Delta(\mathbf{k})$ diverges along nodal directions, then the conditions for nonlocality ($\xi > \lambda$) may be satisfied at low temperatures where the paramagnetic response is dominated by the nodal regions. [119] For a clean d-wave superconductor they predicted that the linear temperature dependence would crossover to a quadratic dependence below $T^*_{NLOC}$,

$$\lambda(T) - \lambda(0) = \frac{\lambda(0)\ln 2}{T_C T^*_{NLOC}} T^2 \tag{48}$$

and

$$T^*_{NLOC} \simeq \Delta(0)\xi(0)/\lambda(0) \tag{49}$$

For YBCO with $T_c \simeq 90$ K this gives $T^*_{NLOC} \simeq 3$ K. The nonlocal correction looks very much like the effect from impurity scattering discussed earlier. As in the impurity scenario, $\lambda(0)$ is also renormalized but the predicted change is only of order 1% since $\lambda(0)$ is determined by the entire Fermi surface, not just the nodal regions. Since the impurity scattering crossover $T^*_{imp}$ in high quality YBCO or BSCCO is of the same order as $T^*_{NLOC}$, the two processes would be difficult to distinguish. However, unlike the effect from impurities, the predicted nonlocality is dependent upon the orientation of the magnetic field and sample boundaries. One must have $H \parallel c-$axis to ensure that the wave-vector defining the spatial variation of the vector potential lies in the conducting plane. Recent penetration depth measurements on $Sr_2RuO_4$ [30] and $CeCoIn_5$ [50] do find experimental evidence for this behavior, as discussed previously and shown in Figure 14.

12.4. Interplane penetration depth

Most unconventional superconductors are highly anisotropic, with normal state conductivity between planes far weaker than in-plane conductivity. Often, the interplane conductivity appears to be incoherent, as evidenced by the vanishing of the Drude peak in the optical conductivity, for example.

Very crudely, incoherent transport occurs once the mean free path for motion normal to the planes becomes smaller than the spacing between them. (The condition for incoherence is a subtle and not fully settled issue.) Meissner screening from interplane supercurrents is therefore weak, as evidenced by an interplane penetration depth, $\lambda_\perp$, that is typically very large compared to the in-plane depth $\lambda_\parallel$. A good example is Bi-2212 where $\lambda_\perp(0) \approx 150 \mu m$.[188] The study of "c-axis" electrodynamics is a large field in which infrared and optical measurements have played a central role. These studies have focused on the behavior of $\lambda_\perp$ with frequency, doping, chemical substitution and the reader is referred to a comprehensive review for further reading. [56, 20] A central result from infrared work has been the demonstration of a strong correlation between $\lambda_\perp$ and the interplane conductivity, $\sigma_\perp$ in cuprates and other unconventional superconductors.[62, 20]

Our discussion will focus on the temperature dependence of $\lambda_\perp$ for which the methods discussed in this paper are more suitable. Certain techniques such as μSR and reversible magnetization are not sensitive to $\lambda_\perp$. Infrared and optical spectroscopy determine $\lambda_\perp$ using $\lambda_\perp(\omega) = 4\pi\omega\sigma_\perp(\omega)/c^2 (\omega \to 0)$ as discussed earlier. In some cases $\lambda_\perp$ is large enough that microwave measurements can determine it from the Josephson plasma frequency using $\omega_p = c(\lambda_\perp \sqrt{\varepsilon})^{-1}$. [146] Here, ε is the dielectric constant at the measurement frequency. Since $\lambda_\perp$ is large, it can often be measured with reasonable (20 %) accuracy from the change in frequency of a resonator, as discussed previously. Scanning SQUID measurements have also been used to determine $\lambda_\perp$ in the organic superconductors κ-(ET)$_2$Cu(NCS)$_2$.[110]

In the simplest approximation one treats the superconductor as a stack of Josephson junctions whose critical current is given by the Ambegaokar-Baratoff relation. [4, 3, 81] For a superconductor with finite gap, $\lambda_\perp(T)$ is then given by,

$$(\rho_s)_\perp \equiv \frac{\lambda_\perp^2(0)}{\lambda_\perp^2(T)} = \frac{\Delta(T)}{\Delta(0)} \tanh\left(\frac{\Delta(T)}{2k_B T}\right)$$

Won and Maki generalized this picture to a d-wave gap function and obtained, [107, 210] a quadratic temperature variation, $1-(\rho_s)_\perp \propto T^2$. More comprehensive theories have taken account of the detailed mechanisms that carry charge from one plane to the next. These include coherent transport, incoherent hopping, elastic impurity scattering and so on. [178, 92, 81] If the transport is purely coherent then $(\rho_s)_\perp$ should have the same temperature dependence as the in-plane superfluid density. Incoherency generally *raises* the exponent of the temperature dependent power law to between n = 2 and n = 3. A large number of measurements on the copper oxides now show $1-(\rho_s)_\perp \propto T^n$ where $2 < n < 2.5$, see Ref. [93] and references therein. Data for YBCO taken by Hoessini et. al. is shown below. *[94]*

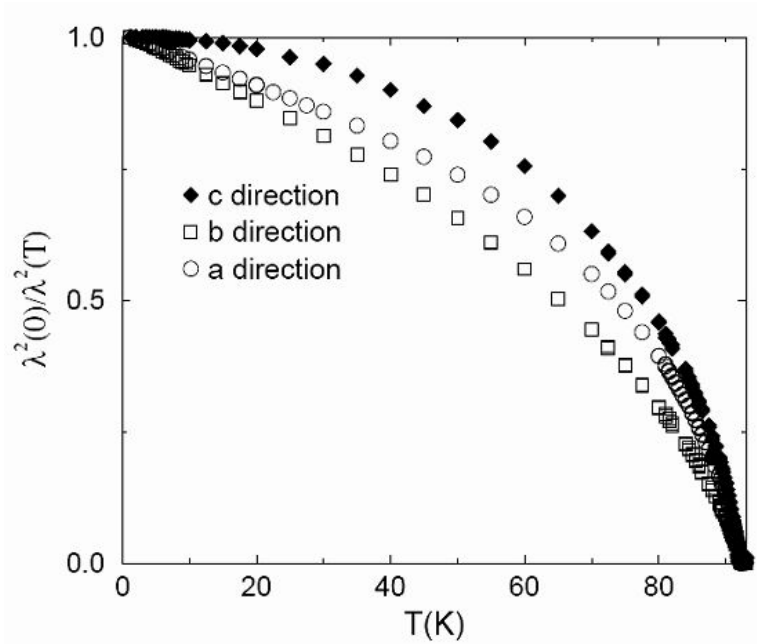

Figure 23. Comparison of c-axis superfluid density and a and b- axis response in YBCO crystal clearly showing the change from linear for a- and b- to quadratic variation the for c-axis. *Reprinted with figure permission from Ref. [94]. Copyright (1998) by the American Physical Society.*

Sheehy *et al.*[187] recently introduced a model that can account for both the temperature and doping dependence of $\lambda_\perp(T)$. In this picture, quasiparticles couple to the field only below a cutoff energy $E_C \propto T_C$. A second parameter characterizes the degree of momentum conservation as electrons hop between planes. Depending upon the relative energy scales, one can obtain exponents ranging from n = 1 to n = 3 for d-wave pairing. This model appears to explain the data on hole doped cuprates, including extremely underdoped YBCO.[93] These same experiments indicate that nodal quasiparticles persist even in extremely underdoped materials and are a robust feature of the copper oxides. [126]

Strong anisotropy does not necessitate incoherent transport. Although $\lambda_\perp/\lambda_\parallel \sim 100$ in the κ-(ET)$_2$ organic superconductors, the interplane transport appears to be nearly coherent. Fig. 21 shows both $\lambda_\perp$ and $(\rho_S)_\perp$ in κ-(ET)$_2$Cu(NCS)$_2$. We find that $1-(\rho_S)_\perp \propto T^{1.2-1.5}$. Fitting the in-plane superfluid density to a pure power law we obtain $1-(\rho_S)_\parallel \propto T^{1.2-1.4}$, with very nearly the same exponent. The uncertainty in the power law exponent comes predominantly from uncertainty in the zero temperature penetration depth. This result would imply coherent interplane transport. Recent magnetoresistance measurements support this conclusion, showing a small but unequivocal three dimensional character to the Fermi surface in this material. [191] It should also be stressed that the power law variation shown in Figure 24 is another demonstration of nodal quasiparticles in the organic superconductors.

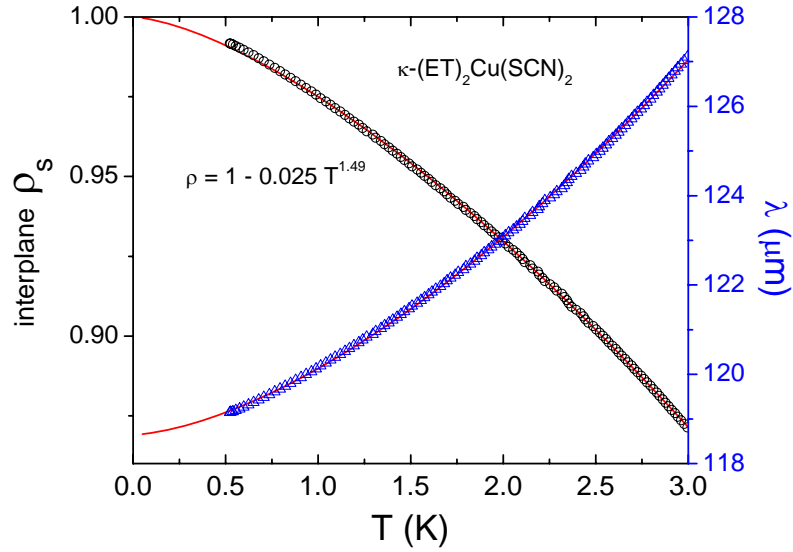

Figure 24. Interplane penetration depth and superfluid density for $\kappa$-(ET)$_2$Cu[SCN]$_2$. Superfluid density is fit to a pure power law.

### 13. Effects of bandstructure: anisotropic and two-gap s-wave superconductors

We now discuss two conventional superconductors, MgB$_2$ and CaAlSi. In addition to being interesting in their own right, these materials serve to illustrate how the formalism described earlier works in more complex situations. Moreover, measurements of the penetration depth in purportedly unconventional materials are sometimes open to alternative interpretations such as strongly anisotropic gaps, multiple gaps, proximity effects and peculiarities of the layer stacking sequence. In such cases one must consider the details of the crystal structure, of the Fermi surface, the density of states as well as intra and interband scattering. In CaAlSi, two 3D bands result in significant interband scattering and a single gap. In MgB$_2$, the difference in dimensionality reduces the interband scattering and two distinct bands (barely) survive, albeit with a single $T_c$.

13.1. MgB$_2$

The discovery of superconductivity at 39 K in the non-cuprate material MgB$_2$ has generated a huge amount of interest. [40, 214, 41, 42] Although there were some early suggestions that the gap function might have nodes, the bulk of evidence now supports a unique and equally interesting situation, namely two-gap superconductivity. This situation is unusual. Most superconductors show multi-band conduction, but due to interband scattering the gap has the same magnitude on all bands. The question of multi-gap superconductivity in *d*-metals was first considered theoretically by Suhl, Matthias and Walker [197] who showed the importance of interband scattering. The Fermi surface of MgB$_2$ is shown in Figure 25.

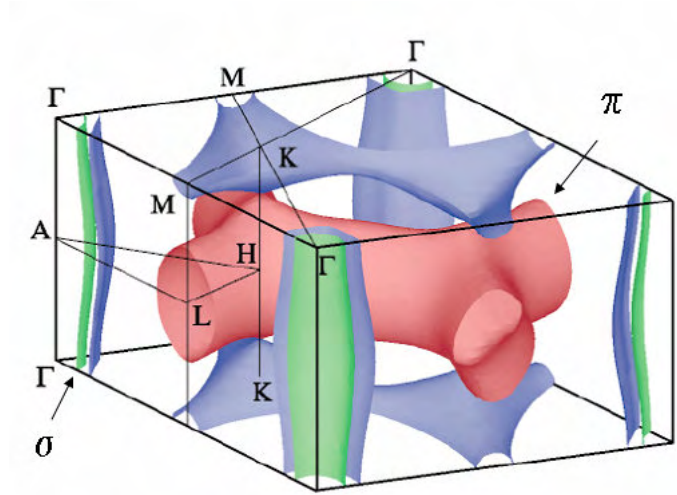

Figure 25. Bandstructure of MgB$_2$ superconductor. Quasi 2D σ band as well as 3D π bands are indicated by arrows. *Reprinted figure with permission from Ref. [117]. Copyright (2001) by the American Physical Society.*

Although strong interband scattering leads to a single $T_c$, the gap magnitudes on the π and σ surfaces are significantly different. This was first observed in tunneling measurements where two conductance peaks were observed, one of which was more easily suppressed in a magnetic field. [186, 218]

Some of the earliest penetration depth measurements on MgB$_2$ wires indicated s-wave pairing but with a minimum gap magnitude 42% of the weak coupling BCS value [167]. Measurements on single crystals by Manzano *et al.* [142] are shown in Figure 26. $\rho_a$ and $\rho_c$ refer to full superfluid densities in the a and c directions respectively. Solid lines through the data are fits using contributions $\rho_\pi$ and $\rho_\sigma$ (also shown) with the "α-model". One considers two distinct gaps on two different bands indexed by $k=1,2$. If $\Delta^k$ does not depend on the wavevector, it can be removed from the integral in Eq. (7). This allows one to define the relative weight of each band to the superfluid density,

$$\rho_{ii}(T) = x_{ii}\rho_1(T) + (1-x_{ii})\rho_2(T) \tag{50}$$

where

$$x_{ii} = \frac{X_{ii}^1}{X_{ii}^1 + X_{ii}^2}, \quad X_{ii}^k = \int \frac{\left(v_F^{ik}\right)^2}{\left|\mathbf{v}_F^k\right|} dS^k \tag{51}$$

[33]. This expression was successfully used to model the penetration depth in MgB$_2$ where two distinct gaps exist. [68]. For the arbitrary case of a highly anisotropic Fermi surface and anisotropic or nodal gap, the complete version of Eq. (7) must be solved.

The large deviation from a single gap model, labelled $\rho_{BCS}$, is clear. It was found that the larger gap is on the σ sheet, $\Delta_\sigma = 75 \pm 5\ K \approx 1.1\Delta_{BCS}$ while smaller gap is on the π sheet, $\Delta_\pi = 29 \pm 2\ K \approx 0.42\ \Delta_{BCS}$. [142] These values are close to those obtained by mutual inductance measurements on MgB$_2$ films [108] and from μSR data. [156] They are also close to the predicted theoretical values. [37] Measurements by Fletcher *et al.* determined the temperature dependence of the anisotropy in the interplane versus in-plane penetration depth. [68]. It was found that at low temperatures $\lambda_c/\lambda_{ab}$ is approximately unity, increasing to about 2 at $T_c$. This is opposite to the temperature dependence of the coherence length anisotropy, which decreases from about 6 to about 2. These various experiments have provided firm quantitative evidence for the two-gap nature of superconductivity in MgB$_2$.

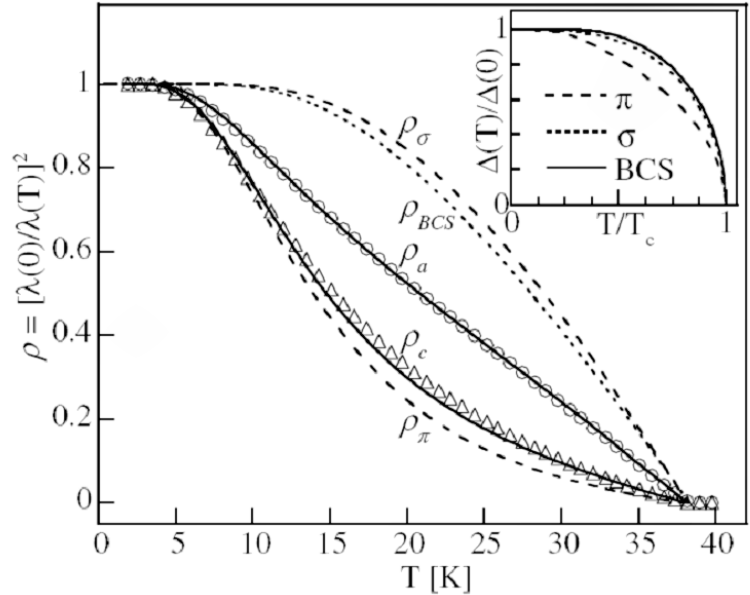

Figure 26. Two components of the superfluid density in single crystal MgB$_2$.. Solid lines are the fits to the α-model, Eq. (50). Long-dashed line are the separate contributions form the σ and π bands, respectively. The short-dashed line is th e weak – coupling BCS. Inset shows temperature dependencies of the two gaps used for the fits. *Reprinted figure with permission from Ref. [68]. Copyright (2005) by the American Physical Society.*

13.2. CaAlSi
It is believed that two distinct gaps survive in MgB$_2$ because of reduced interband scattering due to the different dimensionality of 2D σ and 3D π bands. CaAlSi is isotsructural with MgB$_2$ with a transition temperature ranging from 6.2 K to 7.7 K. Band structure calculations show highly hybridized three-dimensional interlayer and $\pi^*$ bands. [75, 150] Although, most studies of CaAlSi indicate s-wave pairing, deviations from a single isotropic gap behavior have been reported. [74, 199, 135, 97, 96] Magnetic measurements indicate a fully developed s-wave BCS gap. [74, 73, 97] Angle-resolved photoemission spectroscopy [204] revealed the same gap magnitude on the two bands with moderate strong coupling value for the reduced gap, $2\Delta(0)/T_c \approx 4.2$. Together with specific heat measurements [135] it provided reliable evidence for a three-dimensional moderately strong-coupled s-wave BCS superconductivity. On the other hand, μSR studies have been interpreted as evidence of either one highly anisotropic or two distinct energy gaps.[120] Furthermore, 5-fold and 6-fold stacking sequence of (Al,Si) layers corresponding to two different values of $T_c$ were found [182]. Therefore, an experimental study of in- and out-of-plane superfluid density was needed to understand anisotropic superconducting gap structure and mechanism of superconductivity in AlB$_2$ type compounds.

Our penetration depth measurements, shown in Figure 27, indicate that this material is an anisotropic 3D s-wave superconductor. Both components of the superfluid density were measured for both types of CaAlSi (low- and high-$T_c$). To determine the gap anisotropy, the data was fit to a calculation of the superfluid density with an ellipsoidal gap function, $\Delta = \Delta(0)/\sqrt{1-\varepsilon\cos^2(\theta)}$ in the weak coupling BCS limit.

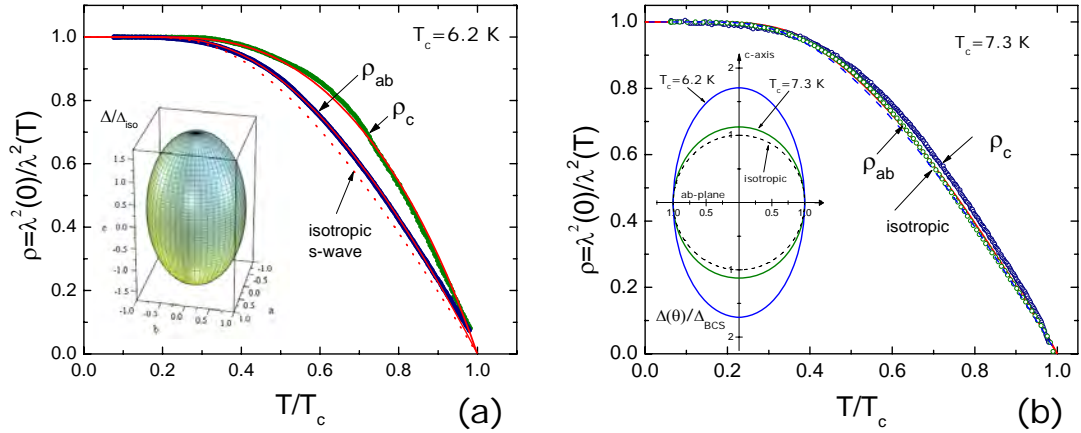

Figure 27. Two components of the superfluid density measured in single crystal CaAlSi. (**left**) lower – $T_c$ sample. Solid lines show results of the full 3D BCS fit to Eqs. (14) and (15) with an ellipsoidal gap as fit parameter, shown in the inset. (**right**) – higher – $T_c$ sample showing almost isotropic response.

Figure 27 shows the result of simultaneously fitting the two components of the superfluid density in a **(a)** lower-$T_c$ and **(b)** higher-$T_c$ sample of CaAlSi. In a higher $T_c$ sample the anisotropy is greatly reduced. [174] For all samples studied, we find that the temperature dependencies of both in-plane and out-of-plane superfluid density are fully consistent with single-gap anisotropic s-wave superconductivity. The gap magnitude in the ab-plane is close to the weak-coupling BCS value while the c-axis values are somewhat larger. Our results suggest that scattering is not responsible for the difference in $T_c$. Scattering would lead to a suppression of the gap anisotropy. A gap with average value $\overline{\Delta}$ and variation $\delta\Delta$ on the Fermi surface can only survive if $\hbar\tau^{-1} \ll \sqrt{\overline{\Delta}\delta\Delta}$, where τ is the impurity scattering rate. [149] However, the values of resistivity are very close for both low- and high - $T_c$ samples, 45 μΩ·cm and 33 μΩ·cm, respectively. The qualitative trend in anisotropy is just the opposite and very pronounced. Also, a 15% suppression of $T_c$ by non-magentic impurities requires very large concentrations. This would significantly smear the transition, which was not observed. It appears, therefore, that the gap anisotropy in CaAlSi abruptly decreases as $T_c$ increases from 6.2 to 7.3 K. A plausible mechanism comes from the analysis of the stacking sequence of (Al/Si) hexagonal layers. [182] There are two structures - 5-fold and 6-fold stacking corresponding to low and higher – $T_c$ samples, respectively. Buckling of (Al,Si) layers is greatly reduced in a 6-fold structure, which leads to the enhancement of the density of states, hence higher $T_c$. Our results suggest that reduced buckling also leads to an almost isotropic gap function. This may be due significant changes in the phonon spectrum and anisotropy of the electron-phonon coupling. To the best of our knowledge, Figure 27 represents the first attempt to use a 3D superconducting gap as fitting parameter to the full 3D BCS problem.

## 14. Magnetic superconductors

The coexistence of magnetism and superconductivity is an exceedingly complex subject, growing larger by the day. We restrict ourselves to one example in which a magnetic ordering transition has a profound influence on the penetration depth: the electron-doped copper oxide SCCO. [173] In the parent compound $Sm_2CuO_4$, rare earth $Sm^{3+}$ ions order at 5.95 K. [102] The ordering is ferromagnetic within each layer parallel to the conducting planes and antiferromagnetic from one layer to the next. [198] Ce doping and subsequent oxygen reduction result in a superconductor $Sm_{1.85}Ce_{0.15}CuO_{4-\delta}$ (SCCO) with $T_c \approx 23 K$.

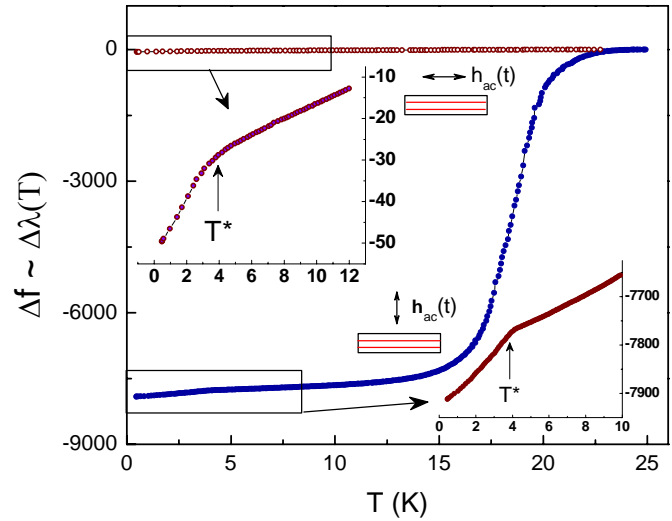

Figure 28. Temperature variation of the penetration depth in SCCO superconductor in two orientations. Insets show low-temperature regions where a distinct break of diamagnetic signal is evident.

Figure 28 shows the penetration depth in single crystal SCCO in for two orientations. The interplane shielding is very weak, yet it shows a transition to more diamagnetic state below about 4 K. The in-plane response shows this effect more clearly.

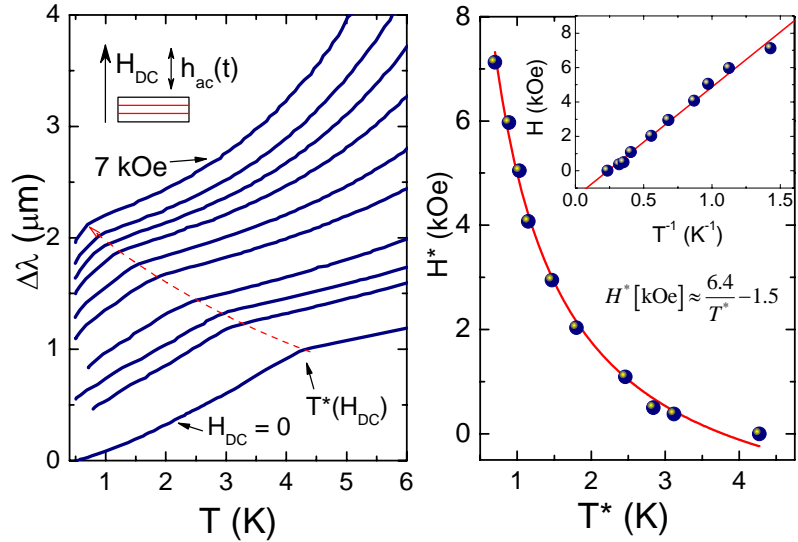

Figure 29. **(left)** Field dependence of the downturn observed in single crystal SCCO. **(right)** the downturn field as function of temperature.

Figure 29 (left) shows the low temperature penetration depth measured in SCCO for several values of a magnetic field applied along the c-axis. First, there is clear evidence for a phase transition near $T^*(H) \approx 4\ K$ that is rapidly suppressed by the magnetic field. Second, the penetration depth drops below the transition indicating stronger diamagnetic screening. This enhanced diamagnetism is quite different from the *weaker* screening that results from paramagnetic impurities discussed earlier. The two effects are consistent with a model involved a spin-freezing transition at $T^*(H)$. Work during the 1980's on the effects on the effect of random impurities on superconductive properties showed that while spin fluctuations will be pairing breaking, the freezing out of these processes will reduce spin-flip scattering and lead to a sharpened density of quasiparticle states. [184] The latter leads, in turn, to stronger diamagnetic screening and a shorter penetration depth.

A spin-glass type of transition is suggested by the rapid suppression of the transition with magnetic field. The Neel temperature of an ordinary antiferromagnet is normally insensitive to fields on this scale. In fact, heat capacity measurements on the parent compound $Sm_2CuO_4$ [69] in fields up to 9 T showed only a tiny shift of the $Sm^{3+}$ Neel temperature. By contrast, the transition temperature in a disordered spin system can be a strong function of magnetic field. In fact, $T^*(H)$ shown in Figure 29 (right) has precisely the same functional form observed for the field dependence of the spin freezing transition in $Fe_{92}Zr_8$, a well-known spin glass. [181] Evidently, the processes required to induce superconductivity in $Sm_2CuO_4$ result in a disordered spin system, although is not clear that the spins actually freezing are $Sm^{3+}$. Spin freezing of $Cu^{2+}$ near 4 K is a well-known phenomenon in other electron-doped copper oxides. [122] It is possible that such a transition is driven by the interaction between $Cu^{2+}$ and $Sm^{3+}$ spins. This example also serves to demonstrate the invaluable role of an external field in interpreting penetration depth measurements in these complex systems.

## 15. Effective penetration depth in the mixed state

Although vortex motion is not the focus of this article, it may have a substantial effect on any measurement in which magnetic fields are present, whether or not vortices are the focus of interest. When vortices are present, the total penetration depth acquires a new contribution. In general this "vortex penetration depth" term can depend upon field, frequency, temperature, orientation and pinning strength. Vortex motion is a complex subject and we will only touch on aspects relating to penetration depth measurements.

A simple model for vortex motion treats the displacement **ū** as a damped harmonic oscillator with a restoring force proportional a constant $\alpha_L$ known as the Labusch parameter, a damping term proportional to the vortex viscosity $\eta = BH_{c2}/\rho_n$, and a driving term due the AC Lorentz force on the vortex, $\sim \vec{j}(t) x \vec{B}$. The inertial term, proportional to vortex mass, is generally ignored. This model was first developed and tested experimentally by Gittleman and Rosenblum [77]. They demonstrated a crossover from pinned flux motion to viscous flux flow as the frequency was increased beyond a "pinning" frequency $\omega_p = \alpha_L/\eta$. Since the advent of high temperature superconductivity, enormous attention has focused on the new features of the H-T phase diagram. [27, 36] It is widely believed that over a substantial portion of the diagram the vortex lattice is melted or at least very weekly pinned, so vortices can be easily displaced. Also, the much higher temperatures that occur in copper oxide superconductors imply that flux flow may be thermally assisted. The effect of this process on the penetration depth was first analyzed by Coffey and Clem [51-53] and by Brandt [35]. A good summary is given in [36]. We briefly describe the model.

A generalized complex penetration depth relevant to small amplitude AC response is given by

$$\lambda^2 = \lambda_L^2 + \frac{B^2}{4\pi}\left(\frac{\alpha_L}{1-i/\omega\tau} + i\omega\eta\right)^{-1} \approx \lambda_L^2 + \lambda_C^2 \frac{1-i/\omega\tau}{1+i\omega\tau_0} \quad (52)$$

$\lambda_L$ is the London penetration depth which, as we have seen, may also depend upon B. $\lambda_C$ is the Campbell pinning length, discussed below. The pinning frequency is defined as $\omega_p \equiv \tau_0^{-1}$ where $\tau_0 = \eta/\alpha_L = B^2/(\rho_{FF}\alpha_L)$. For conventional superconductors the pinning frequency is in the microwave regime but it can be much lower for other materials. $\rho_{FF} = B^2/\eta = \rho_n B/H_{c2}$ is the usual flux flow resistivity. Within the Coffey-Clem model, vortices are also subject to a random force allowing for the possibility of surmounting a field and temperature dependent activation barrier $U(T,B) = U_0(1-T/T_C)^{3/2}/B$. [202] This "thermally-assisted" flux flow [106] is described by an effective time $\tau \approx \tau_0 \exp(U/k_B T)$ which appears in Eq.(52). Since the activation barrier $U(T,B)$ falls with the increase of both *T* and *B*, one may cross from the pinning to flux flow regime by raising either one. The last expression in Eq.(52) is an approximation used when $U \gg k_B T$. In the original model the field dependence of the London penetration depth was taken as,

$$\lambda_L(T,B) = \frac{\lambda_L(T,B=0)}{\sqrt{1-B/B_{c2}}} \quad (53)$$

Eq. (53) is the conventional pair breaking contribution to the field dependence, as discussed earlier. As we have seen, however, the field dependence for *d*-wave superconductors can be much more complex at low temperatures. The Campbell length is given by

$$\lambda_C^2 = \frac{C_{nn}}{\alpha_L} \quad (54)$$

where $\alpha_L$ is the Labusch parameter (note that we use pinning force per volume, not per unit of vortex length),

$$\alpha_L = \frac{j_c B}{c r_p} \quad (55)$$

where $r_p$ is the radius of the pinning potential determined by the maximum pinning force when the vortex is displaced out of equilibrium and $C_{nn}$ is the relevant elastic modulus - $C_{11}$ (compressional modulus) for field parallel to the surface or $C_{44}$ (tilt modulus) for magnetic field perpendicular to the surface. Both modulii are proportional to $B^2$ and therefore,

$$\lambda_C^2 = \frac{B^2}{4\pi\alpha_L} \propto \frac{B}{j_c} \qquad (56)$$

At low frequencies and not too large temperatures and fields, the response is in-phase with the AC field. The effective penetration depth given by

$$\lambda^2 = \lambda_L^2 + \lambda_C^2 \qquad (57)$$

The Campbell contribution, $\lambda_C$, rapidly dominates the London depth, leading to $\lambda(B) \sim \sqrt{B}$. As either the field or temperature is increased, the system crosses over to flux flow, again with an approximate $\lambda(B) \sim \sqrt{B}$ behavior for large fields.

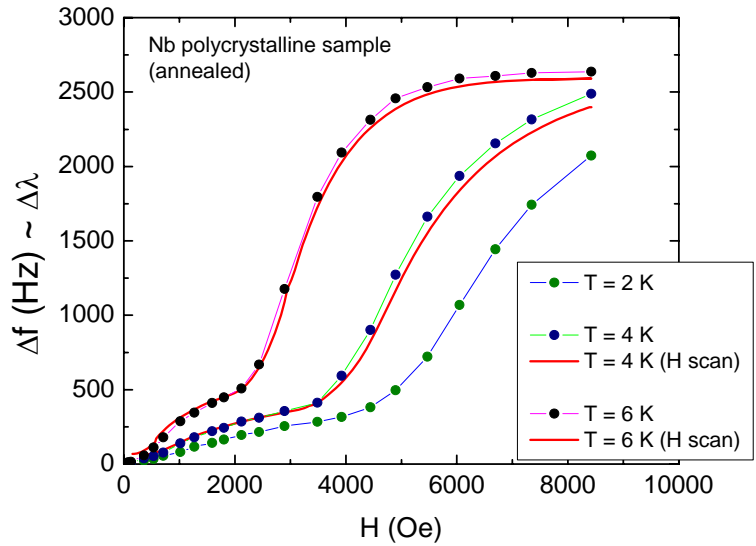

Figure 30. Crossover from pinning to flux-flow regime as a function of magnetic field in polycrystalline Nb sample. The solid lines are field scans at constant temperature. Symbols are obtained from the temperature scans at different fixed fields.

Figure 30 shows the oscillator frequency shift versus field for a stress–free polycrystalline Nb sample. The crossover from pinning to flux flow is seen in both temperature (at different DC fields) and in magnetic field (at different temperatures). Although the Coffey-Clem model was developed in the context of high temperature superconductivity, it is clearly relevant even to conventional materials like Nb. The fact that *T* and *H* scans give identical results implies that the flux profile has no significant inhomogeneity, as required for validity of the model. If a large inhomogeneity in the vortex profile exists, as in the Bean model, the situation is more complicated. The supercurrent that maintains the flux gradient can bias vortices to a different point in the pinning well, leading to a current dependent Labusch parameter. We refer the reader to Ref. [172] for further details.

### 16. Proximity diamagnetism
Among the variety of magnetic phenomena that may affect the penetration depth we discuss one last subject, the proximity effect. As is well known, a superconductor in proximity to a normal metal may induce pairing correlations in the normal metal. [72] For a clean normal metal, these correlations extend a distance $\xi_N = \hbar v_F / 2\pi k_B T$ which can be very large at low temperatures. The normal metal may now carry a quasi-Meissner current and therefore make the combined system appear to have enhanced diamagnetism. Figure 31 shows the effect on the penetration depth**.** [167] The data is shown for a bundle of $180\,\mu m$ diameter MgB$_2$ wires that were coated with a layer of Mg (roughly $2\,\mu m$ thick) leftover from the growth process. Below approximately 5 Kelvin, the penetration depth exhibits a strong negative concavity corresponding to enhanced diamagnetism. In the clean limit, proximity diamagnetism is predicted to turn on at $T \approx 5\,T_{Andreev}$ where the Andreev temperature is defined by $\xi_N(T_{Andreev}) = d$, the film thickness. [65] The diamagnetic downturn vanished completely after dissolving the Mg layer away in alcohol. $\lambda$ for the MgB$_2$ wires left behind exhibited the exponential decrease expected for a superconductor whose minimum energy gap is roughly $0.4\Delta_{BCS}$.

The inset to Figure 31 shows the magnetic field dependence of the proximity diamagnetism. A field of roughly 300 Oe was sufficient to entirely quench the effect. The suppression occurs as the external field exceeds the breakdown field $H_B \sim \phi_0 e^{-d/\xi_N}/d\lambda_N$ of the normal metal film. Here d is the film thickness, $\lambda_N = \sqrt{4\pi e^2/m}$ is (formally) the London penetration depth of the normal metal and $\phi_0$ is the flux quantum. [65] Since the film thickness was not uniform, thicker regions with smaller breakdown fields were quenched first, accounting for the gradual suppression of the diamagnetism shown. The fits shown in the inset to Figure 31 assumed a log-normal distribution of Mg thickness d with the mean and variance of d as free parameters. In addition to being interesting in its own right, proximity diamagnetism is a clear indicator of residual metallic flux that is sometimes left over from the growth process. As Figure 31 shows, its presence can lead to serious errors in interpreting the low temperature behavior of $\lambda(T)$.

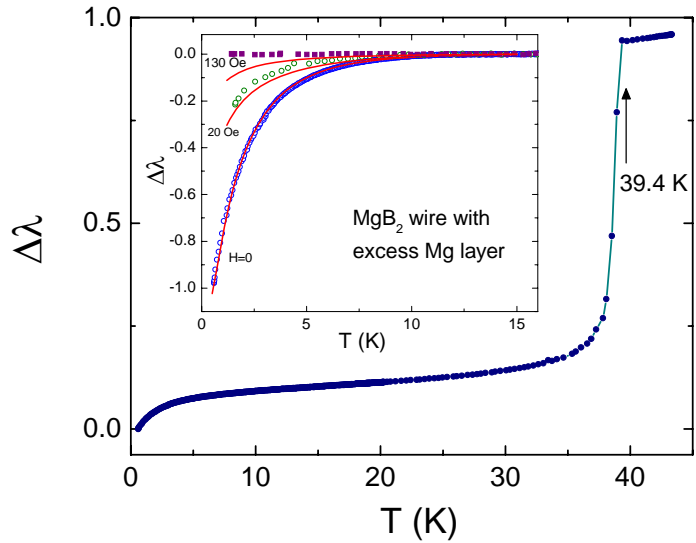

Figure 31. Proximity effect induced superconductivity in MgB$_2$ wires covered by excess Mg layer. Inset shows the low-temperature part. Solid lines are the fits to the dirty proximity effect model.

## 17. Conclusions

Penetration depth measurements were among the earliest to show evidence for nodal quasiparticles in the copper oxide superconductor YBCO. Since that time a sizable number of superconductors have been discovered for which penetration depth measurements indicate some form of unconventional pairing. These materials include hole and electron doped copper oxides, κ-(ET)$_2$X organic superconductors and heavy fermion materials. In general, a reliable determination of the pairing symmetry is extremely difficult and several experimental probes are required before a consensus is reached. We have tried to show how modern penetration depth techniques can contribute to this effort. Many of the technical details required to interpret penetration depth data have been discussed. We have also discussed some unique conventional superconductors to show how penetration depth can be used to explore a complex gap structure on the Fermi surface. The imposition of an external magnetic field has been a particularly useful tool, allowing one to discriminate among a variety of magnetic phenomena including surface Andreev bound states, proximity diamagnetism, vortex motion, magnetic ordering and rare earth paramagnetism. We hope the reader is convinced that penetration depth measurements remain an essential tool in understanding newly discovered superconductors.

## 18. Acknowledgements.

Our special thanks are to Antony Carrington with whom we studied and discussed many of the described effects and to Vladimir G. Kogan and John R. Clem who contributed significantly to understanding of various aspects of this work. Help with data acquisition and analysis by A. Snezhko, T. Olheiser and D. Lawrie is greatly appreciated. We would like to thank Alexey A. Abrikosov, Sergey L. Bud'ko, Paul C. Canfield, Lewis W. Carpenter, Patrick Fournier, Vadim Geshkenbein, Nigel D. Goldenfeld, Alexander A. Gurevich, Peter J. Hirschfeld, Alexey Koshelev, Joerg Schmalian Valerii Vinokur and Taceht Xunil, for numerous useful discussions. Ames Laboratory is operated for the U.S. Department of Energy by Iowa State University under Contract No. W-7405-ENG-82. This work was supported in part by the Director for Energy Research, Office of Basic Energy Sciences. R. P. acknowledges support from NSF Grant # DMR-06-03841 and from the Alfred P. Sloan Research Foundation. R.W.G. acknowledges support from NSF grant # DMR-05-03882.

## References

[1] Abrikosov A A 1988 *Fundamentals of the Theory of Metals, Pt.2*: Elsevier)
[2] Alff L, Meyer S, Kleefisch S, Schoop U, Marx A, Sato H, Naito M and Gross R 1999 Anomalous Low Temperature Behavior of Superconducting Nd1.85Ce0.15CuO4-y *Phys. Rev. Lett.* **83** 2644
[3] Ambegaokar V and Baratoff A 1963 Tunneling Between Superconductors *Phys. Rev. Lett.* **10** 486-9
[4] Ambegaokar V and Baratoff A 1963 Tunneling Between Superconductors (Erratum) *Phys. Rev. Lett.* **11** 104
[5] Anderson P W and Morel P 1961 Generalized Bardeen-Cooper-Schrieffer States and the Proposed Low-Temperature Phase of Liquid He3 *Phys. Rev.* **123** 1911-34
[6] Anlage S, Wu D-H, Mao J, Mao S N, Xi X X, Venkatesan T, Peng J L and Greene R L 1994 Electrodynamics of Nd1.85Ce0.15CuO4: Comparison with Nb and YBa2Cu3O7-δ *Phys. Rev. B* **50** 523
[7] Annett J, Goldenfeld N and Leggett A 1996 Constraints on the pairing state of the cuprate superconductors *Journal of Low Temperature Physics* **105** 473-82
[8] Annett J, Goldenfeld N and Renn S R 1991 Interpretation of the temperature dependence of the electromagnetic penetration depth in YBa2Cu3O7-δ *Phys. Rev. B* **43** 2778-82
[9] Annett J F 1997 Pairing symmetry and pairing interactions in the cuprates *NATO ASI Ser., Ser. 3* **32** 405-14
[10] Annett J F 1999 Unconventional pairing in anomalous superconductors *Physica C* **317-318** 1-8
[11] Annett J F, Goldenfeld N and Renn S R 1990 *Physical Properties of High Temperature Superconductors II,* ed D M Ginsberg (New Jersey: World Scientific) pp 571, Chapter 9

[12] Aprili M, Badica E and Greene L H 1999 Doppler Shift of the Andreev Bound States at the YBCO Surface *Phys. Rev. Lett.* **83** 4630-3

[13] Arai T, Ichimura K, Nomura K, Takasaki S, Yamada J, Nakatsuji S and Anzai H 2001 Tunneling spectroscopy on the organic superconductor κ-(BEDT-TTF)2Cu(NCS)2 using STM *Phys. Rev. B* **63** 104518

[14] Armitage N P, Lu D H, Feng D L, Kim C, Damascelli A, Shen K M, Ronning F, Shen Z-X, Onose Y, Taguchi Y and Tokura Y 2001 Superconducting Gap Anisotropy in Nd1.85Ce0.15CuO4: Results from Photoemission *Phys. Rev. Lett.* **86** 1126-9

[15] Balian R and Werthamer N R 1963 Superconductivity with Pairs in a Relative p Wave *Phys. Rev.* **131** 1553-64

[16] Barash Y S, Kalenkov M S and Kurkijarvi J 2000 Low-temperature magnetic penetration depth in d-wave superconductors: Zero-energy bound state and impurity effects *Physical Review B* **62** 6665-73

[17] Bardeen J, Cooper L N and Schrieffer J R 1957 Theory of superconductivity *Physical Review* **108** 1175-204

[18] Baskaran G 1996 Why is Sr2RuO4 not a high Tc superconductor? Electron correlation, Hund's coupling and p-wave instability *Physica B* **223** 490

[19] Basov D N, Liang R, Bonn D A, Hardy W N, Dabrowski B, Quijada M, Tanner D B, Rice J P, Ginsberg D M and Timusk T 1995 In-Plane Anisotropy of the Penetration Depth in YBa2Cu3O7-x and YBa2Cu4O8 Superconductors *Phys. Rev. Lett.* **74** 598-601

[20] Basov D N and Timusk T 2005 Electrodynamics of high-Tc superconductors *Reviews of Modern Physics* **77** 721

[21] Behnia S, Behnia K and Deluzet A 1998 Heat Conduction in κ-(BEDT-TTF)2Cu(NCS)2 *Phys. Rev. Lett.* **81** 4728

[22] Bhattacharya A, Zutic I, Valls O T, Goldman A M, Welp U and Veal B 1999 Angular Dependence of the Nonlinear Transverse Magnetic Moment of YBa2Cu3O6.95 in the Meissner State *Phys. Rev. Lett.* **82** 3132-5

[23] Bianchi A, Movshovich R, Capan C, Pagliuso P G and Sarrao J L 2003 Possible Fulde-Ferrell-Larkin-Ovchinnikov Superconducting State in CeCoIn[sub 5] *Physical Review Letters* **91** 187004-4

[24] Bickers N E, Scalapino D J and White S R 1989 Conserving Approximations for Strongly Correlated Electron Systems: Bethe-Salpeter Equation and Dynamics for the Two-Dimensional Hubbard Model
 *Phys. Rev. Lett.* **62** 961-4

[25] Bidinosti C P, Hardy W N, Bonn D A and Liang R 1999 Magnetic Field Dependence of λ in YBa2Cu3O6.95: Results as a Function of Temperature and Field Orientation *Phys. Rev. Lett.* **83** 3277-80

[26] Biswas A, Fournier P, Qazilbash M M, Smolyaninova V N, Balci H and Greene R L 2002 Evidence of a d- to s-Wave Pairing Symmetry Transition in the Electron-Doped Cuprate Superconductor Pr2-xCexCuO4 *Phys. Rev. Lett.* **88** 207004

[27] Blatter G, Feigelman M V, Geshkenbein V B, Larkin A I and Vinokur V M 1994 Vortices in high-temperature superconductors *Review of Modern Physics* **66** 1125-388

[28] Blumberg G, Koitzsch A, Gozar A, Dennis B S, Kendziora C A, Fournier P and Greene R L 2002 Nonmonotonic dx2-y2 Superconducting Order Parameter in Nd2-xCexCuO4 *Phys. Rev. Lett.* **88** 107002

[29] Bonalde I, Bramer-Escamilla W and Bauer E 2005 Evidence for Line Nodes in the Superconducting Energy Gap of Noncentrosymmetric CePt3Si from Magnetic Penetration Depth Measurements *Physical Review Letters* **94** 207002/1-/4

[30] Bonalde I, Yanoff B D, Salamon M B, Van Harlingen D J, Chia E M E, Mao Z Q and Maeno Y 2000 Temperature Dependence of the Penetration Depth in Sr2RuO4: Evidence for Nodes in the Gap Function *Physical Review Letters* **85** 4775-8

[31]  Bonalde I, Yanoff B D, Van Harlingen D J, Salamon M B and Maeno Y 2000 Magnetic penetration depth and the symmetry of the order parameter in Sr2RuO4 *Physica C* **341-348** 1695-6

[32]  Bonn D A, Kamal S, Zhang K, Liang R, Baar D J, Klein E and Hardy W N 1994 Comparison of the influence of Ni and Zn impurities on the electromagnetic properties of YBa2Cu3O6.95 *Phys. Rev. B* **50** 4051-63

[33]  Bouquet F, Wang Y, Fisher R A, Hinks D G, Jorgensen J D, Junod A and Phillips N E 2001 Phenomenological two-gap model for the specific heat of MgB2 *Europhys. Lett.* **56** 856-62

[34]  Brandow B 1999 Characteristic features of the exotic superconductors: Evidence for a common pairing mechanism *International Journal of Modern Physics B* **13** 3482-8

[35]  Brandt E H 1991 Penetration of magnetic ac fields into type-II superconductors *Phys. Rev. Lett.* **67** 2219-22

[36]  Brandt E H 1995 The flux-line lattice in superconductors *Reports on Progress in Physics* **58** 1465-594

[37]  Brinkman A, Golubov A A, Rogalla H, Dolgov O V, Kortus J, Kong Y, Jepsen O and Andersen O K 2002 Multiband model for tunneling in MgB2 junctions *Phys. Rev. B* **65** 180517

[38]  Broun D M, Morgan D C, Ormeno R J, Lee S F, Tyler A W, Mackenzie A P and Waldram J R 1997 In-plane microwave conductivity of the single-layer cuprate Tl2Ba2CuO6+δ *Phys. Rev. B* **56** R11443-R6

[39]  Bucholtz L J and Zwicknagl G 1981 Identification of p-wave superconductors *Phys. Rev. B* **23** 5788-96

[40]  Buzea C and Yamashita T 2001 Review of the superconducting properties of MgB<SUB>2</SUB> *Superconductor Science and Technology* **14** R115

[41]  Canfield P C and Bud'ko S L 2002 Magnesium diboride: one year on *Physics World* **15** 29-34

[42]  Canfield P C and Crabtree G W 2003 Magnesium diboride: better late than never *Physics Today* **56** 34-40

[43]  Carbotte J P 1990 Properties of boson-exchange superconductors *Reviews of Modern Physics* **62** 1027-157

[44]  Carbotte J P and Marsiglio F 2003 Electron-phonon superconductivity *Physics of Superconductors* **1** 233-345

[45]  Carrington A, Bonalde I J, Prozorov R, Giannetta R W, Kini A M, Schlueter J, Wang H H, Geiser U and Williams J M 1999 Low-Temperature Penetration Depth of κ-(ET)2Cu[N(CN)2]Br and κ-(ET)2Cu(NCS)2 *Phys. Rev. Lett.* **83** 4172-5

[46]  Carrington A, Giannetta R W, Kim J T and Giapintzakis J 1999 Absence of nonlinear Meissner effect in YBa2Cu3O6.95 *Phys. Rev. B: Condens. Matter Mater. Phys.* **59** R14173-R6

[47]  Carrington A, Manzano F, Prozorov R, Giannetta R W, Kameda N and Tamegai T 2001 Evidence for Surface Andreev Bound States in Cuprate Superconductors from Penetration Depth Measurements *Physical Review Letters* **86** 1074-7

[48]  Chandrasekhar B S and Einzel D 1993 The superconducting penetration depth from the semiclassical model *Annalen der Physik (Berlin, Germany)* **2 (new volume 505)** 535-46

[49]  Chen Q, Kosztin I and Levin K 2000 Unusual Thermodynamical and Transport Signatures of the BCS to Bose-Einstein Crossover Scenario below Tc *Phys. Rev. Lett.* **85** 2801-4

[50]  Chia E E M, Van Harlingen D J, Salamon M B, Yanoff B D, Bonalde I and Sarrao J L 2003 Nonlocality and strong coupling in the heavy fermion superconductor CeCoIn5: A penetration depth study *Physical Review B* **67** 014527/1-/4

[51]  Clem J R and Coffey M W 1991 Radio-frequency surface impedance of type-II superconductors: dependence upon the magnitude and angle of an applied static magnetic field *Physica C: Superconductivity and Its Applications (Amsterdam, Netherlands)* **185-189** 1915-16

[52]  Coffey M W and Clem J R 1991 Unified theory of effects of vortex pinning and flux creep upon the rf surface impedance of type-II superconductors *Phys. Rev. Lett.* **67** 386-9

[53]  Coffey M W and Clem J R 1992 Theory of rf magnetic permeability of isotropic type-II superconductors in a parallel field *Phys. Rev. B* **45** 9872-81
[54]  Coffey T, Bayindir Z, DeCarolis J F, Bennett M, Esper G and Agosta C C 2000 Measuring radio frequency properties of materials in pulsed magnetic fields with a tunnel diode oscillator *Review of Scientific Instruments* **71** 4600-6
[55]  Cooper J R 1996 Power-law dependence of the ab-plane penetration depth in Nd1.85Ce0.15CuO4-y *Physical Review B: Condensed Matter* **54** R3753-R5
[56]  Cooper S L and Gray K E 1994 *Physical Properties of High-Temperature Superconductors,* ed D M Ginsberg (Singapore: World Scientific)
[57]  Curro N J, Simovic B, Hammel P C, Pagliuso P G, Sarrao J L, Thompson J D and Martins G B 2001 Anomalous NMR magnetic shifts in CeCoIn5 *Phys. Rev. B* **64** 180514
[58]  Dagotto E 1994 Correlated electrons in high-temperature superconductors *Rev. Mod. Phys.* **66** 763-840
[59]  Dahm T and Scalapino D J 1999 Nonlinear current response of a d-wave superfluid *Physical Review B: Condensed Matter and Materials Physics* **60** 13125-30
[60]  Degrift C V 1975 Tunnel diode oscillator for 0.001 ppm measurements at low temperatures *Rev. Sci. Instrum.* **46** 599-607
[61]  Desirant M and Shoenberg D 1948 Penetration of magnetic field in superconductors. I. Measurements on thin cylinders *Proceedings of the Physical Society, London* **60** 413-24
[62]  Dordevic S V, Singley E J, Basov D N, Komiya S, Ando Y, Bucher E, Homes C C and Strongin M 2002 Global trends in the interplane penetration depth of layered superconductors *Phys. Rev. B* **65** 134511
[63]  Einzel D, Hirschfeld P J, Gross F, Chandrasekhar B S, Andres K, Ott H R, Beuers J, Fisk Z and Smith J L 1986 Magnetic field penetration depth in the heavy-electron superconductor uranium-beryllium (UBe13) *Physical Review Letters* **56** 2513-16
[64]  Elsinger H, Wosnitza J, Wanka S, Hagel J, Schweitzer D and Strunz W 2000 κ-(BEDT-TTF)2Cu[N(CN)2]Br: A Fully Gapped Strong-Coupling Superconductor *Phys. Rev. Lett. , ()* **84** 6098-101
[65]  Fauchère A L and Blatter G 1997 Magnetic breakdown in a normal-metal-superconductor proximity sandwich *Phys. Rev. B* **56** 14102-7
[66]  Fiory A T, Hebard A F, Mankiewich P M and Howard R E 1988 Penetration depths of high Tc films measured by two-coil mutual inductances *Applied Physics Letters* **52** 2165-7
[67]  Fletcher J D, Carrington A, Piener P, Rodiµere P, Brison J P, Prozorov R, Olheiser T and Giannetta R W 2006 Penetration depth study of anisotropic superconductivity in 2H-NbSe2 *cond-mat*
[68]  Fletcher J D, Carrington A, Taylor O J, Kazakov S M and Karpinski J 2005 Temperature-Dependent Anisotropy of the Penetration Depth and Coherence Length of MgB2 *Physical Review Letters* **95** 097005/1-/4
[69]  Fournier P 2000 private communication.
[70]  Fulde P and Ferrell R A 1964 Superconductivity in a Strong Spin-Exchange Field *Phys. Rev.* **135** A550-A63
[71]  Fulde P, Keller J and Zwicknagl G 1988 Theory of Heavy Fermion systems *Sol. St. Phys.* **41** 1
[72]  Gennes P G d 1966 *Superconductivity of Metals and Alloys* (New York: W. A. Benjamin Inc.)
[73]  Ghosh A K, Hiraoka Y, Tokunaga M and Tamegai T 2003 Anisotropic vortex pinning in the layered intermetallic superconductor CaAlSi *Physical Review B* **68** 134503/1-/6
[74]  Ghosh A K, Tokunaga M and Tamegai T 2003 Angular dependence of the upper critical field in CaAlSi single crystal: Deviation from the Ginzburg-Landau anisotropic mass model *Physical Review B* **68** 054507/1-/5
[75]  Giantomassi M, Boeri L and Bachelet G B 2005 Electrons and phonons in the ternary alloy CaAl[sub 2 - x]Si[sub x] as a function of composition *Physical Review B (Condensed Matter and Materials Physics)* **72** 224512-9

[76]  Ginzburg V L 1952 On account of anisotropy in the theory of superconductivity *Zh. Eksp. Teor. Fiz.* **23** 326
[77]  Gittleman J I and Rosenblum B 1966 Radio-Frequency Resistance in the Mixed State for Subcritical Currents *Phys. Rev. Lett.* **16** 734-6
[78]  Gorter C J and Casimir H 1934 Superconductivity. I *Physica (The Hague)* **1** 306-20
[79]  Gorter C J and Casimir H 1934 The thermodynamics of the superconducting state *Zeitschrift fuer Technische Physik* **15** 539-42
[80]  Gorter C J and Casimir H B G 1934 Zur Thermodynamik des Supraleitenden Zustandes *Phys. Z.* **35** 963
[81]  Graf M J, Palumbo M, Rainer D and Sauls J A 1995 Infrared conductivity in layered d-wave superconductors *Phys. Rev. B* **52** 10588
[82]  Granada C M, da Silva C M and Gomes A A 2002 Magnetic moments of transition impurities in antiperovskites *Solid State Communications* **122** 269-70
[83]  Grosche F M, Julian S R, Walker I R, Freye D M, Haselwimmer R K W and Lonzarich G G 1998 Magnetically mediated superconductivity in heavy fermion compounds *Nature* **394** 39
[84]  Gross-Alltag F, Chandrasekhar B S, Einzel D, Hirschfeld P J and Andres K 1991 London field penetration in heavy fermion superconductors *Zeitschrift fuer Physik B: Condensed Matter* **82** 243-55
[85]  Gross F, Chandrasekhar B S, Einzel D, Andres K, Hirschfeld P J, Ott H R, Beuers J, Fisk Z and Smith J L 1986 Anomalous temperature dependence of the magnetic field penetration depth in superconducting uranium-beryllium(UBe13) *Zeitschrift fuer Physik B: Condensed Matter* **64** 175-88
[86]  Hao Z and Clem J R 1991 Limitations of the London model for the reversible magnetization of type-II superconductors *Phys. Rev. Lett.* **67** 2371–4
[87]  Hardy W N, Bonn D A, Morgan D C, Liang R and Zhang K 1993 Precision measurements of the temperature dependence of l in yttrium barium copper oxide (YBa2Cu3O6.95): strong evidence for nodes in the gap function *Physical Review Letters* **70** 3999-4002
[88]  Hardy W N, Kamal S and Bonn D A 1998 Magnetic penetration depths in cuprates: A short review of measurement techniques and results *NATO ASI Series, Series B: Physics* **371** 373-402
[89]  He T, Huang Q, Ramirez A P, Wang Y, Regan K A, Rogado N, Hayward M A, Haas M K, Slusky J S, Inumara K, Zandbergen H W, Ong N P and Cava R J 2001 Superconductivity in the non-oxide perovskite MgCNi3 *Nature* **411** 54-6
[90]  Hegger H, Petrovic C, Moshopoulou E G, Hundley M F, Sarrao J L, Fisk Z and Thompson J D 84 Pressure-Induced Superconductivity in Quasi-2D CeRhIn5 *Phys. Rev. Lett.* **2000** 4986-9
[91]  Hirschfeld P J and Goldenfeld N 1993 Effect of strong scattering on the low-temperature penetration depth of a d-wave superconductor. *Physical Review B* **48** 4219
[92]  Hirschfeld P J, Quinlan S M and Scalapino D J 1997 c-axis infrared conductivity of a dx2t-y2-wave superconductor with impurity and spin-fluctuation scattering *Phys. Rev. B* **55** 12742
[93]  Hoessini A, Broun D M, Sheehy D E, Davis J P, Franz M, Hardy W N, Liang R and Bonn D A 2004 Survival of the d-Wave Superconducting State near the Edge of Antiferromagnetism in the Cuprate Phase Diagram *Phys. Rev. Lett.* **93** 107003
[94]  Hosseini A, Kamal S, Bonn D A, Liang R and Hardy W N 1998 c-Axis Electrodynamics of YBa2Cu3O7-δ *Phys. Rev. Lett.* **81** 1298-301
[95]  Hu C-R 1994 Midgap surface states as a novel signature for dxa2-xb2-wave superconductivity *Phys. Rev. Lett.* **72** 1526–9
[96]  Imai M, Nishida K, Kimura T and Abe H 2002 Superconductivity of Ca(Al0.5,Si0.5)2, a ternary silicide with the AlB2-type structure *Applied Physics Letters* **80** 1019-21
[97]  Imai M, Sadki E-H S, Abe H, Nishida K, Kimura T, Sato T, Hirata K and Kitazawa H 2003 Superconducting properties of single-crystalline Ca(Al0.5Si0.5)2: a ternary silicide with the AlB2-type structure *Physical Review B* **68** 064512/1-/7

[98]   Ishida K 1998 Spin-triplet superconductivity in Sr2RuO4 identified by 17O Knight shift *Nature* **396** 658
[99]   Ishiguro T, Yamaji K and Saito G 1998 *Organic Superconductivity* (Berlin: Springer-Verlag)
[100]  Izawa K, Yamajuchi H, Sasaki T and Matsuda Y 2002 Superconducting Gap Structure of κ-(BEDT-TTF)2Cu(NCS)2 Probed by Thermal Conductivity Tensor *Phys. Rev. Lett.* **88** 027002
[101]  Jacobs T, Sridhar S, Li Q, Gu G D and Koshizuka N 1995 In-Plane and c-^-Axis Microwave Penetration Depth of Bi2Sr2Ca1Cu2O8+δ Crystals *Phys. Rev. Lett.* **75** 4516-9
[102]  Jiang B, O B-H and Markert J T 1992 Electronic, steric, and dilution effects on the magnetic properties of Sm2-xMxCuO4-y (M=Ce, Y, La, and Sr): Implications for magnetic pair breaking *Phys. Rev. B* **45** 2311-8
[103]  Joynt R and Taillefer L 2002 The superconducting phases of UPt3 *Rev. Mod. Phys.* **74** 235–94
[104]  Karkin A, Goshchitskii B, Kurmaev E, Ren Z A and Che G C 2002 Superconducting properties of atomic-disordered compound MgCNi3 *http://arxiv.org/abs/cond-mat/0209575* 1-6
[105]  Katanin A A 2005 Antiferromagnetic fluctuations, symmetry and shape of the gap function in the electron-doped superconductors: The functional renormalization-group analysis *arXiv:cond-mat/0511120* 1-4
[106]  Kes P H, Aarts J, Berg J v d, Beek C J v d and Mydosh J A 1989 Thermally assisted flux flow at small driving forces *Superconductor Science and Technology* **1** 242
[107]  Khavkine I, Kee H-Y and Maki K 2004 Supercurrent in Nodal Superconductors *Phys. Rev. B* **70** 184521
[108]  Kim M-S, Skinta J A, Lemberger T R, Kang W N, Kim H-J, Choi E-M and Lee S-I 2002 Reflection of a two-gap nature in penetration-depth measurements of MgB2 film *Phys. Rev. B* **66** 064511
[109]  Kim M S, Skinta J A, Lemberger T R, Tsukada A and Naito M 2003 Magnetic Penetration Depth Measurements of Pr2–xCexCuO4– Films on Buffered Substrates: Evidence for a Nodeless Gap *Phys. Rev. Lett.* **91** 087001
[110]  Kirtley J A, Moler K A, Schlueter J A and Williams J M 1999 Inhomogeneous interlayer Josephson coupling in kappa -(BEDT-TTF)2Cu(NCS)2 *J. Phys. Condens. Matter* **11** 2007
[111]  Klemm R A 1998 What is the symmetry of the high-Tc order parameter? *Int. J. Mod. Phys. B* **12** 2920-31
[112]  Kogan V G 1981 London approach to anisotropic type-II superconductors *Phys. Rev. B* **24** 1572–5
[113]  Kogan V G, Prozorov R, Bud'ko S L, Canfield P C, Thompson J R and Karpinsky J 2006 Effect of field dependent core size on magnetization of high κ superconductors *to be published*
[114]  Kokales J D, Fournier P, Mercaldo L V, Talanov V V, Greene R L and Anlage S M 2000 Microwave Electrodynamics of Electron-Doped Cuprate Superconductors *Phys. Rev. Lett.* **85** 3696
[115]  Kondo H and Moriya T 1998 Spin Fluctuation-Induced Superconductivity in Organic Compounds *J. Phys. Soc. Jpn.* **67** 3695-8
[116]  Kondo T 2006 The study of the electronic structure of (Bi,Pb)2(Sr,La)2CuO6+d high temperature superconductor by Angle Resolved Photoemission Spectroscopy (ARPES). *private communication*
[117]  Kortus J, Mazin I I, Belashchenko K D, Antropov V P and Boyer L L 2001 Superconductivity of Metallic Boron in MgB2 *Phys. Rev. Lett.* **86** 4656-9
[118]  Kosztin I, Chen Q, Jankó B and Levin K 1998 Relationship between the pseudo- and superconducting gaps: Effects of residual pairing correlations below Tc *Phys. Rev. B* **58** R5936-R9
[119]  Kosztin I and Leggett A J 1997 Nonlocal Effects on the Magnetic Penetration Depth in d-Wave Superconductors *Phys. Rev. Lett.* **79** 135-8
[120]  Kuroiwa S, Takagiwa H, Yamazawa M, Akimitsu J, Ohishi K, Koda A, Higemoto W and Kadono R 2004 Unconventional behavior of field-induced quasiparticle excitation in

Ca(Al0.5Si0.5)2 *Journal of the Physical Society of Japan* **73** 2631-4
[121] Larkin A I and Ovchinnikov Y N 1964 Nonuniform state of superconductors *Zh. Eksp. Teor. Fiz.* **47** 1136-46
[122] Lascialfari A, Ghigna P and Tedoldi F 2003 Magnetic phase diagram of Nd1.85Ce0.15CuO4+δ from magnetization and muon spin relaxation measurements *Phys. Rev. B* **68** 104524
[123] Le L P, Luke G M, Sternlieb B J, Wu W D, Uemura Y J, Brewer J H, Riseman T M, Stronach C E, Saito G, Yamochi H, Wang H H, Kini A M, Carlson K D and Williams J M 1992 Muon-spin-relaxation measurements of magnetic penetration depth in organic superconductors (BEDT-TTF)2-X: X=Cu(NCS)2 and Cu[N(CN)2]Br *Phys. Rev. Lett.* **68** 1923-6
[124] Lee I J, Naughton M J, Danner G M and Chaikin P M 1997 Anisotropy of the Upper Critical Field in (TMTSF)2PF6 *Phys. Rev. Lett.* **78** 3555-8
[125] Lee J Y, Paget K M, Lemberger T R, Foltyn S R and Wu X 1994 Crossover in temperature dependence of penetration depth λ(T) in superconducting YBa2Cu3O7-δ films *Phys. Rev. B* **50** 3337-41
[126] Lee P A and Wen X-G 1997 Unusual Superconducting State of Underdoped Cuprates *Phys. Rev. Lett.* **78** 4111-4
[127] Lee S-F, Morgan D C, Ormeno R J, Broun D M, Doyle R A, Waldram J R and Kadowaki K 1996 a-b Plane Microwave Surface Impedance of a High-Quality Bi2Sr2CaCu2O8 Single Crystal *Phys. Rev. Lett.* **77** 735-8
[128] Leggett A J 1975 A theoretical description of the new phases of liquid 3He *Rev. Mod. Phys.* **47** 331-414
[129] Lewis H W 1956 Two-Fluid Model of an "Energy-Gap" Superconductor *Phys. Rev.* **102** 1508-11
[130] Li M R, Hirschfeld P J and Wolfle P 1998 Is the Nonlinear Meissner Effect Unobservable? *Physical Review Letters* **81** 5640-3
[131] Lin J and Millis A J 2005 Theory of low-temperature Hall effect in electron-doped cuprates *Physical Review B* **72** 214506/1-/9
[132] Lin J Y, Ho P L, Huang H L, Lin P H, Zhang Y L, Yu R C, Jin C Q and Yang H D 2003 BCS-like superconductivity in MgCNi3 *Physical Review B* **67** 052501/1-/4
[133] Liu C S, Luo H G, Wu W C and Xiang T 2005 Theory of Raman scattering on electron-doped high-Tc superconductor *Los Alamos National Laboratory, Preprint Archive, Condensed Matter* 1-4, arXiv:cond-mat/0512173
[134] London F and London H 1935 The electromagnetic equations of the supraconductor *Proc. R. Soc. London, Ser. A* **149** 71
[135] Lorenz B, Cmaidalka J, Meng R L and Chu C W 2003 Thermodynamic properties and pressure effect on the superconductivity in CaAlSi and SrAlSi *Physical Review B* **68** 014512/1-/6
[136] Louati R, Charfi-Kaddour S, Ali A B, Bennaceur R and Héritier M 2000 Normal-state properties of BEDT-TTF compounds and the superconductivity pairing mechanism *Phys. Rev. B* **62** 5957-64
[137] Mackenzie A P and Maeno Y 2000 p-wave superconductivity *Physica B* **280** 148
[138] Maeda A and Hanaguri T 1998 Magnetic field dependence of the surface impedance in superconductors *Supercond. Rev.* **3** 1-49
[139] Maeno Y, Hashimoto H, Yoshida K, Nishizaki S, Fujita T, Bednorz J G and Lichtenberg F 1994 Superconductivity in a layered perovskite without copper *Nature* **372** 532
[140] Maki K 1998 Introduction to d-wave superconductivity *AIP Conf. Proc.* **438** 83-128
[141] Mansky P A, Chaikin P M and Haddon R C 1994 Evidence for Josephson vortices in (BEDT-TTF)2Cu(NCS)2 *Phys. Rev. B* **50** 15929-44
[142] Manzano F, Carrington A, Hussey N E, Lee S, Yamamoto A and Tajima S 2002 Exponential Temperature Dependence of the Penetration Depth in Single Crystal MgB2 *Physical Review Letters* **88** 047002/1-/4
[143] Mao J, Wu D H, Peng J L, Greene R L and Anlage S M 1995 Anisotropic surface impedance of

YBa2Cu3O7-δ single crystals *Phys. Rev. B* **51** 3316-9
[144] Mao Z Q, Rosario M M, Nelson K D, Wu K, Deac I G, Schiffer P, Liu Y, He T, Regan K A and Cava R J 2003 Experimental determination of superconducting parameters for the intermetallic perovskite superconductor MgCNi3 *Physical Review B* **67** 094502/1-/6
[145] Martin C, Agosta C C, Tozer S W, Radovan H A, Palm E C, Murphy T P and Sarrao J L 2005 Evidence for the Fulde-Ferrell-Larkin-Ovchinnikov state in CeCoIn[sub 5] from penetration depth measurements *Physical Review B (Condensed Matter and Materials Physics)* **71** 020503-4
[146] Matsuda Y, Ong N P, Yan Y F, Harris J M and Peterson J B 1994 Vortex viscosity in YBa2Cu3O7-δ at low temperatures *Phys. Rev. B* **49** 4380-3
[147] Matsui H, Terashima K, Sato T, Takahashi T, Fujita M and Yamada K 2005 Direct Observation of a Nonmonotonic dx2-y2-Wave Superconducting Gap in the Electron-Doped High-Tc Superconductor Pr0.89LaCe0.11CuO4 *Physical Review Letters* **95** 017003/1-/4
[148] Mayaffre H, Wzietek P, Jérome D, Lenoir C and Batail P 1995 Superconducting State of κ-(ET)2CU[N(CN)2]Br Studied by 13C NMR: Evidence for Vortex-Core-Induced Nuclear Relaxation and Unconventional Pairing *Phys. Rev. Lett.* **75** 4122-5
[149] Mazin I I, Andersen O K, Jepsen O, Golubov A A, Dolgov O V and Kortus J 2004 Comment on ``First-principles calculation of the superconducting transition in MgB[sub 2] within the anisotropic Eliashberg formalism'' *Physical Review B* **69** 056501-3
[150] Mazin I I and Papaconstantopoulos D A 2004 Electronic structure and superconductivity of CaAlSi and SrAlSi *Physical Review B* **69** 180512/1-/4
[151] McKenzie R H 1998 *Comments Condens. Matt. Phys.* **18** 309-26
[152] Meissner W and Ochsenfeld R 1933 Ein neuer Effekt bei Eintritt der Supraleitfahigkeit *Die Naturwissenschaften* **21** 787
[153] Meissner W and Ochsenfeld R 1983 A new effect concerning the onset of Superconductivity (English translation by A. M. Forrest) *Eur. J. Phys.* **4** 117-20
[154] Monthoux P, Balatsky A V and Pines D 1991 Toward a theory of high-temperature superconductivity in the antiferromagnetically correlated cuprate oxides *Phys. Rev. Lett.* **67** 3448-51
[155] Nelson K D, Mao Z Q, Maeno Y and Liu Y 2004 Odd-Parity Superconductivity in Sr2RuO4 *Science* **306** 1151 - 4
[156] Niedermayer C, Bernhard C, Holden T, Kremer R K and Ahn K 2002 Muon spin relaxation study of the magnetic penetration depth in MgB2 *Phys. Rev. B* **65** 094512
[157] Oates D E, Park S H and Koren G 2004 Observation of the nonlinear meissner effect in YBCO thin films: evidence for a D-wave order parameter in the bulk of the cuprate superconductors *Physical Review Letters* **93** 197001
[158] Osheroff D D, Richardson R C and Lee D M 1972 Evidence for a New Phase of Solid He3 *Phys. Rev. Lett.* **28** 885-8
[159] Özcan S, Broun D M, Morgan B, Haselwimmer R K W, Sarrao J L, Kamal S, Bidinosti C P and P. J. Turner M 2003 London penetration depth measurements of the heavy-fermion superconductor  near a magnetic quantum critical point *Europhys. Lett.* **62** 412-8
[160] Panagopoulos C, Cooper J R, Peacock G B, Gameson I, Edwards P P, Schmidbauer W and Hodby J W 1996 Anisotropic magnetic penetration depth of grain-aligned HgBa2Ca2Cu3O8+δ *Phys. Rev. B* **53** R2999-R3002
[161] Panagopoulos C, Cooper J R, Xiang T, Peacock G B, Gameson I and Edwards P P 1997 Probing the Order Parameter and the c-Axis Coupling of High- Tc Cuprates by Penetration Depth Measurements *Phys. Rev. Lett.* **79** 2320-3
[162] Panagopoulos C and Xiang T 1998 Relationship between the Superconducting Energy Gap and the Critical Temperature in High- Tc Superconductors *Phys. Rev. Lett. ,  ()* **81** 2336-9
[163] Pippard A B 1953 An experimental and theoretical study of the relation between magnetic field and current in a superconductor *Proc. Roy. Soc. (London)* **A216** 547-68

[164] Porch A, Cooper J R, Zheng D N, Waldram J R, Campbell A M and Freeman P A 1993 Temperature dependent magnetic penetration depth of Co and Zn doped YBa2Cu3O7 obtained from the AC susceptibility of magnetically aligned powders *Physica C* **214** 350

[165] Preosti G, Kim H and Muzikar P 1994 Density of states in unconventional superconductors: Impurity-scattering effects *Phys. Rev. B* **50** 1259-63

[166] Printeric M, Tomic S, Prester M, Drobac D and Maki K 2002 Influence of internal disorder on the superconducting state in the organic layered superconductor κ-(BEDT-TTF)2Cu[N(CN)2]Br *Phys. Rev. B* **66** 174521

[167] Prozorov R, Giannetta R W, Bud'ko S L and Canfield P C 2001 Energy gap and proximity effect in MgB2 superconducting wires *Physical Review B* **64** 180501/1-/4

[168] Prozorov R, Giannetta R W, Carrington A and Araujo-Moreira F M 2000 Meissner-London state in superconductors of rectangular cross section in a perpendicular magnetic field *Physical Review B* **62** 115-8

[169] Prozorov R, Giannetta R W, Carrington A, Fournier P, Greene R L, Guptasarma P, Hinks D G and Banks A R 2000 Measurements of the absolute value of the penetration depth in high-Tc superconductors using a low-Tc superconductive coating *Applied Physics Letters* **77** 4202-4

[170] Prozorov R, Giannetta R W, Fournier P and Greene R L 2000 Evidence for nodal quasiparticles in electron-doped cuprates from penetration depth measurements *Physical Review Letters* **85** 3700-3

[171] Prozorov R, Giannetta R W, Fournier P and Greene R L 2000 Magnetic penetration depth in electron-doped cuprates - evidence for gap nodes *Physica C: Superconductivity and Its Applications (Amsterdam)* **341-348** 1703-4

[172] Prozorov R, Giannetta R W, Kameda N, Tamegai T, Schlueter J A and Fournier P 2003 Campbell penetration depth of a superconductor in the critical state *Physical Review B* **67** 184501/1-/4

[173] Prozorov R, Lawrie D D, Hetel I, Fournier P and Giannetta R W 2004 Field-Dependent Diamagnetic Transition in Magnetic Superconductor Sm1.85Ce0.15CuO4-y *Physical Review Letters* **93** 147001/1-/4

[174] Prozorov R, Olheiser T A, Giannetta R W, Uozato K and Tamegai T 2006 Anisotropic s-wave superconductivity in single crystals CaAlSi from penetration depth measurements *submitted*

[175] Prozorov R, Snezhko A, He T and Cava R J 2003 Evidence for unconventional superconductivity in the nonoxide perovskite MgCNi3 from penetration depth measurements *Physical Review B* **68** 180502/1-/4

[176] Prusseit W, Walter H, Semerad R, Kinder H, Assmann W, Huber H, Kabius B, Burkhardt H, Rainer D and Sauls J A 1999 Observation of paramagnetic Meissner currents—evidence for Andreev bound surface states *Physica C* **318** 396-402

[177] Qazilbash M M, Koitzsch A, Dennis B S, Gozar A, Balci H, Kendziora C A, Greene R L and Blumberg G 2005 Evolution of superconductivity in electron-doped cuprates: Magneto-Raman spectroscopy *Physical Review B* **72** 214510/1-/12

[178] Radtke R J, Kostur V N and K.Levin 1996 Theory of the c-axis penetration depth in the cuprates *Phys. Rev. B* **53** R522-5

[179] Rice T M and Sigrist M 1995 *J. Phys. Condes. Matter* **7** L643

[180] Rosner H, Weht R, Johannes M D, Pickett W E and Tosatti E 2002 Superconductivity near Ferromagnetism in MgCNi3 *Physical Review Letters* **88** 027001/1-/4

[181] Ryan D H, Lierop J v, Pumarol M E, Roseman M and Cadogan J M 2001 Field dependence of the transverse spin freezing transition *Phys. Rev. B* **63** 140405

[182] Sagayama H, Wakabayashi Y, Sawa H, Kamiyama T, Hoshikawa A, Harjo S, Uozato K, Ghosh A K, Tokunaga M and Tamegai T 2006 Two Types of Multistack Structures in MgB2-Type Superconductor CaAlSi *Journal of the Physical Society of Japan (Letters)* **75** 043713

[183] Sauls J A 1994 The order parameter for the superconducting phases of UPt3 *Advances in Physics (Publisher: Taylor & Francis)* **43** 113 - 41

[184] Schachinger E, Stephan W and Carbotte J P 1988 Quasiparticle density of states for a spin-glass superconductor *Phys. Rev. B* **37** 5003-9
[185] Schmalian J 1998 Pairing due to Spin Fluctuations in Layered Organic Superconductors *Phys. Rev. Lett.* **81** 4232-5
[186] Schmidt H, Zasadzinski J F, Gray K E and Hinks D G 2002 Evidence for Two-Band Superconductivity from Break-Junction Tunneling on MgB2 *Phys. Rev. Lett.* **88** 127002
[187] Sheehy D E, Davis T P and Franz M 2004 Unified theory of the ab-plane and c-axis penetration depths of underdoped cuprates *Phys. Rev. B* **70** 054510
[188] Shovkun D V and al. e 2000 Anisotropy of conductivity in high-Tc single crystals *JETP Lett.* **71** 92
[189] Signore P J C, Andraka B, Meisel M W, Brown S E, Z. Fisk A L G, and J. L. Smith , Gross-Alltag F, Schuberth E A and Menovsky A A 1995 Inductive measurements of UPt3 in the superconducting state *Phys. Rev. B* **52** 4446-61
[190] Singer P M, Imai T, He T, Hayward M A and Cava R J 2001 C13 NMR investigation of the superconductor MgCNi3 up to 800 K *Physical Review Letters* **87** 257601/1-/4
[191] Singleton J 2006 APS March meeting. In: *APS March Meeting,* (Baltimore, MD.
[192] Skalski S, Betbeder-Matibet O and Weiss P R 1964 Properties of Superconducting Alloys Containing Paramagnetic Impurities *Phys. Rev.* **136** A1500-A18
[193] Snezhko A, Prozorov R, Lawrie D D, Giannetta R W, Gauthier J, Renaud J and Fournier P 2004 Nodal Order Parameter in Electron-Doped Pr2-xCexCuO4-d Superconducting Films *Physical Review Letters* **92** 157005/1-/4
[194] Sonier J E, Brewer J H and Kiefl R F 2000 μSR studies of the vortex state in type-II superconductors *Rev. Mod. Phys.* **72** 769–811
[195] Soto S M D, Slichter C P, Kini A M, Wang H H, Geiser U and Williams J M 1995 13C NMR studies of the normal and superconducting states of the organic superconductor κ-(ET)2Cu[N(CN)2]Br *Phys. Rev. B* **52** 10364-8
[196] Sridhar S, Wu D-H and Kennedy W 1989 Temperature dependence of electrodynamic properties of YBa2Cu3Oy crystals *Phys. Rev. Lett.* **63** 1873-6
[197] Suhl H, Matthias B T and Walker L R 1959 Bardeen-Cooper-Schrieffer Theory of Superconductivity in the Case of Overlapping Bands *Phys. Rev. Lett.* **3** 552-4
[198] Sumarlin I W, Lynn J W, Neumann D A, Rush J J, Loong C K, Peng J L and Li Z Y 1993 Phonon density of states in rare earth(R) copper oxide (R2CuO4) and superconducting rare earth(R) cerium copper oxide (R1.85Ce0.15CuO4) (R = neodymium, praseodymium) *Physical Review B* **48** 473-82
[199] Tamegai T, Uozato K, Ghosh A K and Tokunaga M 2004 Anisotropic superconducting properties of CaAlSi and CaGaSi *International Journal of Modern Physics B* **19** 369-74
[200] Thouless D J 1960 Perturbation theory in statistical mechanics and the theory of superconductivity *Ann. Phys.* **10** 553
[201] Timusk T and Statt B 1999 The pseudogap in high-temperature superconductors: an experimental survey *Rep. Prog. Phys.* **62** 61-122
[202] Tinkham M 1988 Resistive Transition of High-Temperature Superconductors *Phys. Rev. Lett.* **61** 1658-61
[203] Tinkham M 1996 *Introduction to Superconductivity* (New York: McGraw-Hill Book Co.)
[204] Tsuda S, Yokoya T, Shin S, Imai M and Hase I 2004 Identical superconducting gap on different Fermi surfaces of Ca(Al0.5Si0.5)2 with the AlB2 structure *Physical Review B* **69** 100506/1-/4
[205] Tsuei C C and J.R.Kirtley 2000 Phase-Sensitive Evidence for d-Wave Pairing Symmetry in Electron-Doped Cuprate Superconductors *Phys. Rev. Lett.* **85** 182
[206] Tsuei C C and Kirtley J R 2000 Pairing symmetry in cuprate superconductors *Reviews of Modern Physics* **72** 969-1016
[207] Van Harlingen D J 1995 Phase-sensitive tests of the symmetry of the pairing state in the high-temperature superconductors--Evidence for dx2-y2 symmetry *Reviews of Modern Physics* **67**

515-35
[208] Voelker K and Sigrist M 2002 Unconventional superconductivity in MgCNi3 *http://arxiv.org/abs/cond-mat/0208367* 1-4
[209] Waldram J R, Porch A and Cheah H M 1994 Measurement of the microwave conductivities of high-Tc superconducting powders *Physica C* **232** 189
[210] Won H, Jang H and Maki K 1999 Paramagnetic state in d-wave Superconductors *cond-mat/9901252*
[211] Won H, Jang H, Parker D, Haas S and Maki K 2004 Panorama of Nodal Superconductors *cond-mat/0405099*
[212] Won H and Maki K 2000 Possible f-wave superconductivity in Sr2RuO4? *Europhys. Lett.* **52** 427
[213] Xu D, Yip S K and Sauls J A 1995 Nonlinear Meissner effect in unconventional superconductors *Phys. Rev. B* **51** 16233–53
[214] Yanson I K and Naidyuk Y G 2004 Advances in point-contact spectroscopy: two-band superconductor MgB2 (A review) *Low Temp. Phys.* **30** 261
[215] Yip S K and Sauls J A 1992 Nonlinear Meissner effect in CuO superconductors *Phys. Rev. Lett.* **69** 2264-7
[216] Yoshimura H and Dai S H 2004 Angular dependence of the superconducting order parameter in electron-doped high-temperature superconductors *Journal of the Physical Society of Japan* **73** 2057-60
[217] Yu W, Liang B and Greene R L 2005 Magnetic-field dependence of the low-temperature specific heat of the electron-doped superconductor Pr1.85Ce0.15CuO4 *Physical Review B: Condensed Matter and Materials Physics* **72** 212512/1-/3
[218] Zalk M v, Brinkman A, Golubov A A, Hilgenkamp H, Kim T H, Moodera J S and Rogalla H 2006 Fabrication of multiband MgB2 tunnel junctions for transport measurements *Superconductor Science and Technology* **19** S226
[219] Žutić I and Valls O T 1998 Low-frequency nonlinear magnetic response of an unconventional superconductor *Phys. Rev. B* **58** 8738-48